\documentclass[]{article}
\pdfoutput=1

\newcommand\affil[1]{$^{#1}$\ignorespaces}
\newcommand\affiliation[2]{\vskip-.5\parskip\relax{\centering{\footnotesize
$^{#1}$#2\relax}\vskip-\parskip}}

\usepackage[a4paper, includehead, includefoot, nomarginpar, margin=1.1in ]{geometry}

\usepackage{fancyhdr}

\usepackage{soul}
\usepackage{graphicx}
  \setkeys{Gin}{draft=false}
\usepackage{setspace}
\usepackage{amsmath}
\usepackage{amssymb}
\usepackage{url}

\RequirePackage{apacite}
\let\cite\shortcite %xx So get et al. with three authors the first time.
\let\citet\shortciteA %xx Ditto.
\RequirePackage{url}
\AtBeginDocument{%
}

\usepackage{fancyhdr}
\renewcommand{\headrulewidth}{0pt}
\usepackage[small]{titlesec}

\usepackage[title]{appendix}
\usepackage{caption}
\title{ General Local Reactive Boundary Condition for Dissolution and Precipitation  Using the Lattice Boltzmann Method \\ \vspace{1em}}
\author{Julius Weinmiller\affil{1,2,}\thanks{ Corresponding author: julius.weinmiller@dlr.de}\ ,
     Martin P. Lautenschlaeger\affil{1,2},
     Benjamin Kellers\affil{1,2}, \\ 
     Timo Danner\affil{1,2,}\thanks{ Co-corresponding author: timo.danner@dlr.de}\ ,
     Arnulf Latz\affil{1,2,3}}

\date{}

\begin{document}

\fancyhead{} % clear all header fields
\fancyhead[R]{\small \textit{General Local RBC for Dissolution and Precipitation using LBM}}
\fancyfoot{} % clear all footer fields
\fancyfoot[R]{\thepage}

\fancypagestyle{plain}{%
  \fancyhf{} 
  \fancyfoot[l]{\textit{Preprint submitted to arXiv}}
  \fancyfoot[c]{}
  \fancyfoot[r]{\textit{June 16, 2023}}
  \renewcommand{\headrulewidth}{0pt}
}

\maketitle

%% ------------------------------------------------------------------------ %%
%
%  AUTHORS AND AFFILIATIONS
%
%% ------------------------------------------------------------------------ %%

\vspace{-1em}
\affiliation{1}{German Aerospace Center (DLR), Institute of Engineering Thermodynamics, 89081 Ulm, Germany}
\affiliation{2}{Helmholtz Institute Ulm for Electrochemical Energy Storage (HIU), 89081 Ulm, Germany}
\affiliation{3}{Ulm University (UUlm), Institute of Electrochemistry, 89081 Ulm, Germany}

\vspace{3em}

%% ------------------------------------------------------------------------ %%
%
%  TEXT
%
%% ------------------------------------------------------------------------ %%

\hrule
\begin{abstract}
\normalsize
A general and local reactive boundary condition (RBC) for studying first-order equilibrium reactions using the lattice Boltzmann method (LBM) is presented. Its main characteristics are accurate reproduction of wall diffusion, invariance to the wall and grid orientation, and absence of nonphysical artefacts. The scheme is successfully tested for different benchmark cases considering diffusion, advection, and reactions of fluids at solid-liquid interfaces. Unlike other comparable RBCs from the literature, the novel scheme is valid for a large range of Péclet and Damköhler numbers, and shows realistic pattern formation during precipitation. In addition, quantitative results are in good accordance with analytical solutions and values from literature. Combining the new RBC with the rest fraction method, Péclet-Reynolds ratios of up to 1000 can be achieved. Overall, the novel RBC accurately models first-order reactions, is applicable for complex geometries, and allows efficiently simulating dissolution and precipitation phenomena in fluids at the pore scale.
\end{abstract}
\vspace{1em}
\hrule
\vspace{1em}
\vspace{1em}
\hrule

\vspace{2em}
\newpage

\section{Introduction}
%{
Dissolution and precipitation of solid solutes occur in a broad range of natural phenomena and technological applications. Typical examples are dissolution of minerals in subsurface hydrology \cite{prikryl_mineral_2017, baqer_review_2022, andrews_fracture_2023}, biofilm growth in nutrient rich environments \cite{jung_pore-scale_2021}, etching into substrates \cite{Cui2019}, and conversion of active material in energy storage systems \cite{fang_pore-scale_2021, Danner2019}. 

Despite their omnipresence, a detailed experimental analysis of such processes is often lacking due to the dominant small length and short time scales on which they occur. As such, computational methods and more specifically continuum approaches are used to describe these phenomena at large. But even they are stretched to their limits, when dissolution and precipitation occur in structurally complex porous media where the relevant physics happen at the pore scale.  

A computational method that has proven to give insight into such mesoscopic phenomena is the lattice Boltzmann method (LBM). It can be applied to predict flow, transport, and reactions in porous media \cite{lautenschlaeger_lattice_2023, guiltinan_twophase_2021, liu_porescale_2021}. Especially over the last decade, LBM has gained importance both technically and application-wise. On the one hand, the ease of meshing and parallelization makes LBM extremely favorable for high-performance computing \cite{latt_palabos_2021, krause_openlbopen_2021, kellers_proceedings_2023}. On the other hand, it can be applied to a broad variety of application fields. In the context of porous media these are for example: oil and gas flow in underground formations \cite{ren_lattice_2015, li_pore-scale_2015}, oil recovery with in-situ combustion \cite{lei_pore-scale_2022}, pore structure evolution in cement manufacturing \cite{Patel2021b,Patel2014}, combustion of porous solid rocket fuel \cite{Wang2018}, water transport and reactions in fuel cells \cite{sarkezi-selsky_lattice_2022}, multi-phase flow in batteries \cite{lautenschlaeger_understanding_2022,lautenschlaeger_homogenized_2022,danner_characterization_2016}, and dissolution and precipitation reactions in stone formations \cite{Chen2014d, matthias_ehrhardt_lattice_2013, tian_lattice_2017, zhang_porescale_2019}. 

In most of these cases, dissolution and precipitation play an important role. Using LBM, such reactions are typically modelled as heterogeneous reactive boundary conditions (RBC). Different approaches to translate macroscopic reaction behavior into mesoscopic schemes are known from literature. They range from simple modified bounce-back schemes \cite{Kang2002, kang_improved_2007}, over pseudo-homogeneous reactions \cite{Patel2014}, and schemes based on interpolation using the wall normal \cite{walsh_interpolated_2010, BatteryRelevance, xie_simple_2021, D2Q5LLi2017}, up to general local boundary schemes for first-order \cite{Patel2016, Verhaeghe2006, Ju2020} and higher-order reactions \cite{kashani_non-linear_2022, Hiorth2013}. Two important scheme characteristics are \textit{generality} and \textit{locality}. Here, \textit{general} means that the underlying scheme is the same regardless of the surrounding cells, i.e.\ corners, edges and flat boundaries use the same scheme. Whereas \textit{local} means that only information of the current boundary cell is used. In combination, they are a necessary basis for versatile and computationally efficient RBCs applicable to complex geometries.

Several approaches have been considered to derive such general local boundary schemes. Three of which are discussed in more detail in this paper: 
1) \citeA{Verhaeghe2006} was amongst the first to develop such a model for multi-component LBM using the momentum transfer analysis of \citeA{bouzidi_momentum_2001}. 
2) \citeA{Patel2016} developed an approach that bases on bouncing back the non-equilibrium part and including the flux as a concentration gradient while correctly capturing the macroscopic wall diffusion. 
3) \citeA{Ju2020} extended an approach, which was originally developed for finite differences \cite{zhang_general_2012} and then converted to a local scheme \cite{meng_boundary_2016}, to also consider the correct macroscopic wall diffusion. 
All aforementioned RBC schemes show robust behavior in cases where the RBC normal is aligned with the lattice. However, cases different from these benchmarks are either missing in the corresponding papers or they show a nonphysical behavior of the corresponding scheme there. 

Therefore, in this paper, a novel RBC scheme is presented which is based on the work of \citeA{Verhaeghe2006}, \citeA{Ju2020}, and \citeA{Patel2016}. It is shown to overcome most deficiencies of these methods. It captures the correct wall diffusion and wall normal behavior, and is applicable to a broad range of reaction regimes. It is shown that the new RBC scheme can be combined with the rest fraction method \cite{sullivan_lattice_2005} to enable numerically stable simulations even at large ratios between the solute diffusivity and solvent viscosity.

The new RBC scheme as well as the RBC schemes of \citeA{Verhaeghe2006}, \citeA{Ju2020}, and \citeA{Patel2016} are tested using numerous benchmark cases that vary in complexity. For the sake of comparability, all schemes are reformulated to a common notation. The test cases include: 1) A robust reaction-diffusion verification case for which analytical solutions exists. 2) Pattern formation in a precipitation process. 3) A sophisticated benchmark case from the literature \cite{Molins2020}, considering reactions in channel flow. All simulations are conducted in 2D. They can, however, be easily extended to 3D.

The paper is organized as follows: In Section~\ref{sec:NumericalMethods}, LBM basics, the rest fraction method, as well as the coupling of LBM to dissolution and precipitation are introduced. This section also covers the reformulation of the aforementioned RBC schemes. The verification and validation of the different schemes are compared and discussed in Section~\ref{sec:Results}. Finally, the findings are summarized and conclusions are given in Section~\ref{sec:Conclusion}.
%}

\section{Numerical Methods}  \label{sec:NumericalMethods}
All LBM simulations presented in this work were conducted using an extended version of the LBM simulation package Palabos \cite{latt_palabos_2021}. There, the rest fraction method, reactive boundary conditions, and dissolution and precipitation were implemented.

\subsection{LBM Fundamentals} \label{sec:LBM}
%{
For classical fluid flow, LBM solves the Boltzmann equation discretized in the phase-space mostly on a uniform grid. The velocity space is represented by the so-called lattice. It consists of a set of velocities $\lbrace\mathbf{e}_i \rbrace$, weights $\lbrace\mathrm{w}_i\rbrace$ and a specific lattice speed of sound $c_{\mathrm{s}}$. It forms the framework to distribute a set of particles that are represented by a particle distribution function, typically called population. In the following, it is indicated by either $f_i$ or $g_i$, where $i$ is the lattice direction. 
Fluid flow emerges through local collision and subsequent streaming of the populations. For reactive flows, the hydrodynamic carrier fluid and the advected scalar field are distinguished using separate lattices and equations. They are coupled by the fluid velocity, determined from the carrier fluid, which enters the advected scalar field. 
%}

\subsubsection*{Hydrodynamic Equations}
%{
The dynamics of the carrier fluid described by the Navier-Stokes (NS) equations are solved using the lattice Boltzmann equation combined with the well-known BGK collision operator \cite{kruger_lattice_2017} 
\begin{equation}
\label{eq:fluid_BGK}
    f_i ( \mathbf{x}+\mathbf{e}_i \Delta t, t+\Delta t ) = f_i ( \mathbf{x}, t) - \frac{1}{\tau_{\mathrm{NS}}} \left( f_i ( \mathbf{x}, t) - f_i^{\mathrm{eq}} ( \mathbf{x}, t)  \right),
\end{equation}
where $\tau_{\mathrm{NS}}$ is the relaxation time which is determined by the fluid viscosity $\nu = c_{\mathrm{s}}^2 \left( \tau_{\mathrm{NS}} - 0.5 \right)$ and $f_i^{\mathrm{eq}}$ is the equilibrium population 
\begin{equation}
\label{eq:fluid_Equilib}
    f_{i}^{\mathrm{eq}}(\rho,\mathbf{u}) = \mathrm{w}_{i}\rho\left( 1 + \frac{\mathbf{e}_{i} \cdot \mathbf{u}}{c_{s}^{2}} + \frac{ \left(\mathbf{u}\cdot\mathbf{e}_i \right)^2 }{2c_{s}^{4}} - \frac{\mathbf{u}\cdot\mathbf{u}}{2c_{s}^{2}}\right).
\end{equation}
The fluid density $\rho$ and the fluid velocity $\mathbf{u}$ are determined from $f_i$ as
\begin{equation}
\label{eq:fluid_BGK_visc}
   \rho = \sum_i f_i\,\,, \qquad \rho \mathbf{u} = \sum_i f_i \mathbf{e}_i.
\end{equation}
In the following, the D2Q9 lattice is applied to solve the hydrodynamic equations in 2D. Further details are given in Appendix~\ref{sec:App_VelocitySet}.

At high relaxation times, the BGK collision operator introduces numerical slip at bounce-back walls leading to inaccuracies and instabilities. To overcome these artefacts, the two relaxation time (TRT) collision operator \cite{TRTGinzburg2008a,TRTHumieres2009} is used in such cases. It is given as
\begin{eqnarray}
\label{eq:fluid_TRT}
    f_i ( \mathbf{x}+\mathbf{e}_i \Delta t, t+\Delta t ) = f_i ( \mathbf{x}, t) \hspace{-1em}&- \frac{1}{\tau_{\mathrm{NS}}^+} \left( f_i^+ ( \mathbf{x}, t) - f_i^{\mathrm{eq+}} ( \mathbf{x}, t)  \right) \nonumber \\ &- \frac{1}{\tau_{\mathrm{NS}}^-} \left( f_i^- ( \mathbf{x}, t) - f_i^{\mathrm{eq-}} ( \mathbf{x}, t)  \right);
\end{eqnarray}
\begin{equation}
\label{eq:fluid_TRT_terms}
    f_i^\pm = \frac{f_i \pm f_{\bar{i}} }{2}, \qquad f_i^\mathrm{eq\pm} = \frac{f_i^\mathrm{eq} \pm f^\mathrm{eq}_{\bar{i}}}{2}.
\end{equation}
Here, $\bar{i}$ is the direction opposite to $i$, i.e.\ $\mathbf{e}_{\bar{i}}=-\mathbf{e}_{i}$. The values of $\tau_{\mathrm{NS}}^+$ and $\tau_{\mathrm{NS}}^-$ are defined as 
\begin{equation}
\label{eq:fluid_TRT_visc}
   \nu = c_{\mathrm{s}}^2 \left( \tau_{\mathrm{NS}}^+ - \frac{1}{2} \right), \qquad \qquad \Lambda = \left( \tau_{\mathrm{NS}}^+ - \frac{1}{2} \right)\left( \tau_{\mathrm{NS}}^- - \frac{1}{2} \right)
\end{equation}
from the viscosity of the fluid $\nu$ and the so-called ``magic'' parameter $\Lambda$.
In the following, $\Lambda=\frac{3}{16}$, if not specified otherwise. It ensures that walls implemented via the bounce-back scheme are located exactly halfway between a solid and a fluid node.
%}

\subsubsection*{Advection-Diffusion Equations}
%{
Advection and diffusion (AD) of a scalar field are described in a similar manner as the hydrodynamic equations given by Eq.~(\ref{eq:fluid_BGK}). In the following, AD populations are indicated by $g_i$.
\begin{equation}
\label{eq:solute_BGK}
    g_i ( \mathbf{x}+\mathbf{e}_i \Delta t, t+\Delta t ) = g_i ( \mathbf{x}, t) - \frac{1}{\tau_{\mathrm{AD}}} \left( g_i ( \mathbf{x}, t) - g_i^{\mathrm{eq}} ( \mathbf{x}, t)  \right).
\end{equation}
Here, $\tau_{\mathrm{AD}}$ is determined by the diffusivity of the scalar $\mathrm{D} = c_{\mathrm{s}}^2 \left( \tau_{\mathrm{AD}} - 0.5 \right)$ and $g_i^{\mathrm{eq}}$ is
\begin{equation}
    \label{eq:solute_equilibrium_function}
    g_i^{\mathrm{eq}} (C, \mathbf{u}) = C \mathrm{w}_i \left[ 1 + \frac{\mathbf{u}\cdot\mathbf{e}_i}{c_{\mathrm{s}}^2} \right].
\end{equation}
Here, $\mathbf{u}$ is the local advection velocity of the corresponding fluid field (cf.\ Eq.~(\ref{eq:fluid_BGK_visc})). From Eq.~(\ref{eq:solute_BGK}), the concentration $C$ of the scalar field is determined as
\begin{equation}
\label{eq:solute_diffu}
    C = \sum_i g_i.
\end{equation}
Note, that using this local advection velocity, the first moment of $g_i$, i.e.\ momentum, is not conserved. Therefore, Eq.~(\ref{eq:solute_BGK}) solves only the AD equation but not the NS equations (cf.\ Eq.~(\ref{eq:fluid_BGK})). With fewer degrees of freedom, it is sufficient to use a reduced D2Q5 lattice (cf.\ Appendix~\ref{sec:App_VelocitySet}) and a linear equilibrium function (cf.\ Eq.(\ref{eq:solute_equilibrium_function})) to solve Eq.~(\ref{eq:solute_BGK}) which significantly decreases computational efforts.

The scalar field can also be solved using the TRT collision operator, for both accuracy and stability reasons. The corresponding set of equations reads
\begin{eqnarray}
\label{eq:solute_TRT}
    g_i ( \mathbf{x}+\mathbf{e}_i \Delta t, t+\Delta t ) = g_i ( \mathbf{x}, t)  \hspace{-1em}&- \frac{1}{\tau_{\mathrm{AD}}^-} \left( g_i^+ ( \mathbf{x}, t) - g_i^{\mathrm{eq+}} ( \mathbf{x}, t)  \right) \nonumber \\ &- \frac{1}{\tau_{\mathrm{AD}}^+} \left( g_i^- ( \mathbf{x}, t) - g_i^{\mathrm{eq-}} ( \mathbf{x}, t)  \right)
\end{eqnarray}
% the swapping of +/- at the tau is NOT a mistake
\begin{equation}
\label{eq:solute_TRT_terms}
    g_i^\pm = \frac{g_i \pm g_{\bar{i}} }{2}, \qquad g_i^\mathrm{eq\pm} = \frac{g_i^\mathrm{eq} \pm g^\mathrm{eq}_{\bar{i}}}{2}.
\end{equation}
Analogous to the hydrodynamic equations, $\tau_{\mathrm{AD}}^+$ and $\tau_{\mathrm{AD}}^-$ depend on both the diffusivity and the magic parameter, $\Lambda=\frac{3}{16}$ by default:
\begin{equation}
\label{eq:solute_TRT_visc}
   D = c_{\mathrm{s}}^2 \left( \tau_{\mathrm{AD}}^+ - \frac{1}{2} \right), \qquad \qquad \Lambda = \left( \tau_{\mathrm{AD}}^+ - \frac{1}{2} \right)\left( \tau_{\mathrm{AD}}^- - \frac{1}{2} \right).
\end{equation}
%}

\subsection{Rest Fraction Method} \label{sec:RestFraction}
%{
Coupling NS and AD via the advection velocity requires the same spatial and temporal discretization of both fields. Together with the two facts that 1) $\tau_j$ should be within the range of $\tau_j\approx [0.6, 1.4]$ to ensure numerical stability and accuracy \cite[p.188]{he_analytic_1997, kruger_lattice_2017} and 2) $\tau_j$ is proportional to $\nu$ or $\mathrm{D}$ (and therefore also to the Reynolds or Péclet number) this strongly limits the range of applicable Reynolds-to-Péclet ratios.

Using the TRT method where $\tau_j\approx [0.6, 4.0]$ can slightly extend this range. However, a technique to significantly increase the range is the rest fraction method. It decouples the discretization and relaxation time by redefining the equilibrium function of the scalar field \cite{sullivan_simulation_2005, sullivan_lattice_2005}. Following \cite{looije_introducing_2018}, here, the original formulation is interpreted as a method to change $c_{\mathrm{s}}^2$ as well as the weights $\mathrm{w}_i$ by the rest fraction $J_0$
\begin{equation}
\label{eq:RestFraction}
    \mathrm{w}_{i}(J_0) = \left\{ \begin{array}{ll} J_0 &  \mathrm{if}\quad i=0, \\
                        c_{\mathrm{s}}^2 / 2 & \mathrm{otherwise}; \end{array} \right.
\end{equation}
\begin{equation}
\label{eq:RestFractionCs2}
    c_{\mathrm{s}}^2 = \frac{1-J_0}{\mathrm{dim}},
\end{equation}
where ``$\mathrm{dim}$'' is 2 for D2Q5 and 3 for D3Q7. This approach does not affect the other equations, i.e.\ Eqs.~(\ref{eq:fluid_BGK})$-$(\ref{eq:solute_TRT_visc}), unlike the original formulation. Note that setting the free parameter $J_0 = 1/3$ for D2Q5 or $J_0 = 1/4$ for D3Q7 results in the commonly used values for $c_{\mathrm{s}}^2$ and $\mathrm{w}_i$.

The rest fraction method allows for tuning the relation between diffusivity and relaxation time, with\ $\mathrm{D} = \frac{(1 - J_0)}{\mathrm{dim}} \left( \tau_{\mathrm{AD}} - 0.5 \right)$. Therefore, Reynolds-to-Péclet ratios of up to 1000 can be achieved by adjusting solely $J_0$, while keeping $\tau \approx 1$.
%}

\subsection{Reactive Boundary Conditions} \label{sec:BoundaryCondition}
%{
All RBC schemes discussed in this paper macroscopically describe a first-order equilibrium reaction given as  
\begin{equation}
    \label{eq:reactionRateBoundary}
    J_{\mathrm{R}} = k_{\mathrm{r}} \left( C_\mathrm{eq} - C_\mathrm{wall}  \right).
\end{equation}
Here, $J_{\mathrm{R}}$ is the reaction rate, interpretable as a population flux. $k_{\mathrm{r}}$ is the reaction rate coefficient and $C_\mathrm{eq}$ and $C_\mathrm{wall}$ are the equilibrium concentration and concentration at the wall, respectively. The AD scalar field can also represent temperature, in which case the population flux $J_{\mathrm{R}}$ describes a heat flux, instead of a reaction.

The macroscopic flux $J_{\mathrm{R}}$ cannot be directly used in LBM due to its mesoscopic character, where the flux is determined using the populations. In addition, $C_\mathrm{wall}$ is unknown and needs to be estimated using an appropriate mesoscopic scheme. Details of how to transform the macroscopic first-order equilibrium reaction (cf.\ Eq.~(\ref{eq:reactionRateBoundary})) into a mesoscopic boundary condition (cf.\ Eq.~(\ref{eq:general_RBC})) are given in the supporting information (SI) Text~S1.1.

\begin{figure}[ht!]
    \centering
    % \captionsetup{width=.7\linewidth}
    \includegraphics[trim={1cm 9.cm 1cm 0cm}, clip, width=0.7\linewidth]{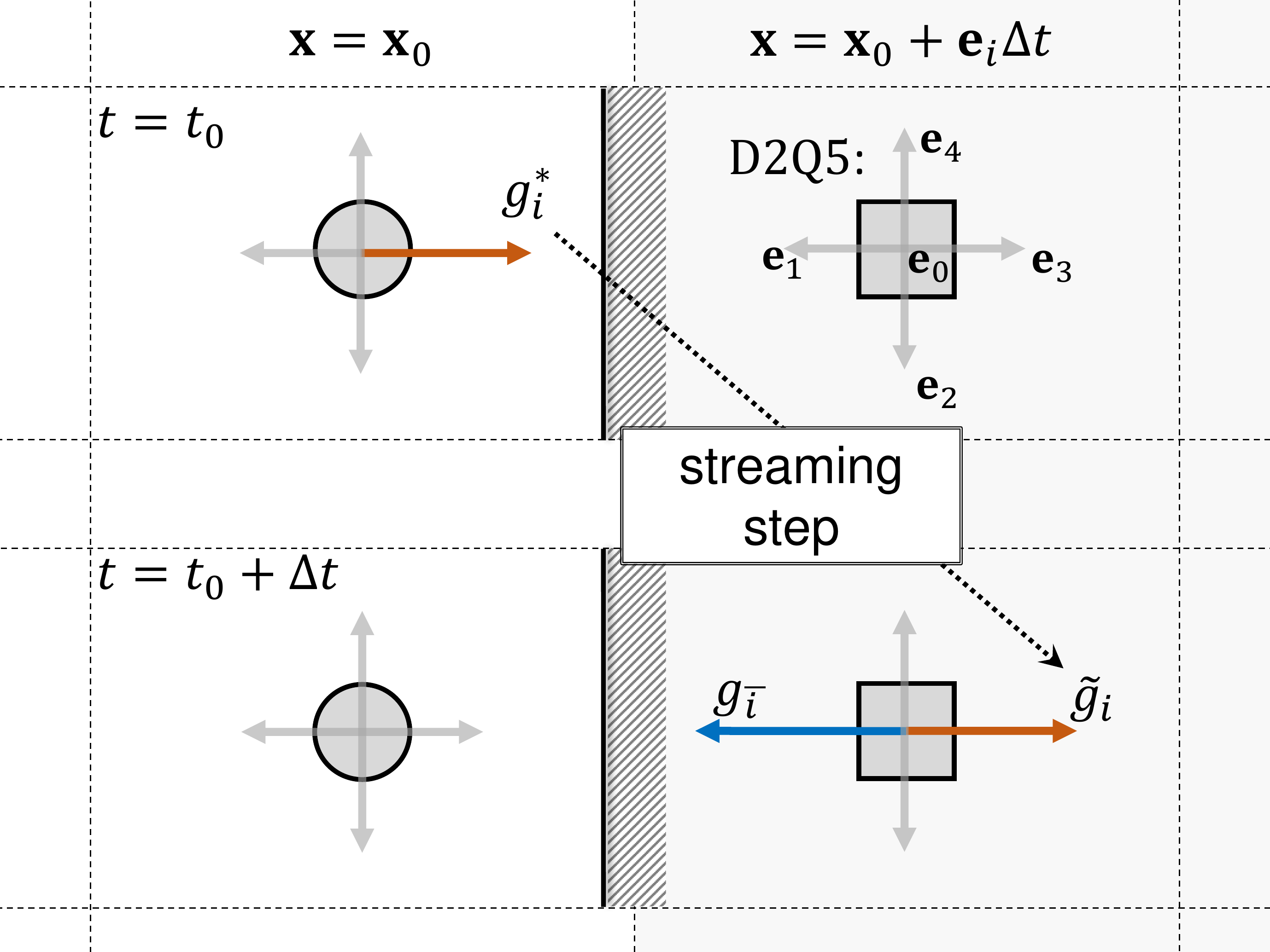}
    \caption{Schematic representation of the first-order RBC. Fluid nodes (circles) are depicted on the left side. Solid nodes (squares) are depicted on the right side. The top and the bottom rows only differ in time, but not in space. The streaming of population $g_i^*~=~g_i( \mathbf{x},\ t_0)$ to $\tilde g_i~=~g_i(\mathbf{x}+\mathbf{e}_i\Delta t,\ t_0+\Delta t)$ are indicated by the orange arrows; the result of the RBC $g_{\bar i}~=~g_{\bar i}(\mathbf{x}+\mathbf{e}_i\Delta t,\ t_0+\Delta t)$ with the blue arrow. In addition, the top right quadrant shows the D2Q5 velocity set ($\mathbf{e}_i$ for $i\in \{0,1,2,3,4\}$).}
    \label{fig:SchmaticBoundaryCondition}
\end{figure}

Fig.~\ref{fig:SchmaticBoundaryCondition} schematically shows the functioning principle of the set of RBC schemes discussed in this paper. It starts with the population $g_i^* = g_i( \mathbf{x},\ t_0)$ pointing towards the wall and before streaming. After streaming this population becomes $\tilde g_i = g_i(\mathbf{x}+\mathbf{e}_i\Delta t,\ t_0+\Delta t)$. The actual main step of the RBC scheme determines the unknown population $g_{\bar i}$ from the known $\tilde g_i$. For the next streaming step $g_{\bar i}$ is required, and it is the final outcome of an RBC scheme.

The RBC schemes of \citeA{Verhaeghe2006}, \citeA{Patel2016} and \citeA{Ju2020}, and the new RBC scheme are presented in the following. For brevity, in the following only the first authors are mentioned and the corresponding schemes are abbreviated to S.V., S.P.J. and S.New, respectively. For comparability they are reformulated to a similar notation. 

It is shown in detail in the SI (cf.\ Text~S1) that they can be expressed in the form
 % \ref{sec:Appendix_BoundariesExpression}
\begin{equation}
    \label{eq:general_RBC}
    g_{\bar i} = \frac{k_i}{1+k_i} 2 \mathrm{w}_i C_{\mathrm{eq}} + \frac{1-k_i}{1+k_i} \tilde{g}_i.
\end{equation}
The difference of the schemes only enters by the definition of the term $k_i$
\begin{eqnarray}
    \label{eq:k_i_S_Ve}
     k_{i}^{(\mathrm{S.V.})} &= c_{\mathrm{s}}^{-2} \ \  k_{\mathrm{r}} \cdot (\mathbf{e}_{\bar i}\cdot\mathbf{n}) \\
    \label{eq:k_i_S2}
     k_{i}^{(\mathrm{S.P.J.})} &= c_{\mathrm{s}}^{-2} \gamma k_{\mathrm{r}} /\ (\mathbf{e}_{\bar i}\cdot\mathbf{n}) \\
    \label{eq:k_i_S_New}
     k_{i}^{(\mathrm{S.New})} &= c_{\mathrm{s}}^{-2} \gamma k_{\mathrm{r}} \cdot (\mathbf{e}_{\bar i}\cdot\mathbf{n}),
\end{eqnarray}
and more specifically by the incorporation of the wall normal term $(\mathbf{e}_{\bar i}\cdot\mathbf{n})$ and the correction factor $\gamma = \tau_{\mathrm{AD}} / (\tau_{\mathrm{AD}} - 0.5)$. This reformulation approach also allows cross-implementation of already developed features from one RBC to the other, e.g.\ moving walls. Here, it is used to highlight the differences of the schemes.

In the scheme of Verhaeghe the wall normal term is included in the numerator and $\gamma$ is not considered. In the schemes of Patel and Ju the wall normal term is included in the denominator and $\gamma$ is considered. Compared to the scheme of Verhaeghe, in the schemes of Patel and Ju the $\gamma$ can be interpreted as a shift in the effective reaction rate. The novel scheme proposed here is only slightly different and combines elements from both the scheme of Verhaeghe and the schemes of Patel and Ju. This, however, significantly improves both accuracy and applicability for complex geometries, and overcomes the $\tau$-dependent wall diffusion as is shown in Section~\ref{sec:Results}.

For all three schemes, the limits of the reaction rate are similar \cite{Verhaeghe2006,Ju2020,Patel2016}: At high reactivity $(k_{\mathrm{r}} \rightarrow \infty)$, the schemes simplify to the anti bounce-back scheme (anti-BB) $g_{\bar i} = 2 \mathrm{w}_i C_{\mathrm{eq}} - \tilde{g}_i$, which sets the wall concentration to the equilibrium concentration. If no reactions occur $(k_{\mathrm{r}}=0)$, the schemes simplify to the common bounce-back method, i.e.\ $g_{\bar i} = \tilde{g}_i$. 
%}

\subsection{Dissolution and Precipitation} \label{sec:DissolutionPrecipitation}
%{
Finally, simulating phase changes such as dissolution and precipitation at solid surfaces is built up on the RBC scheme. After the determination of the reaction flux using the RBC scheme, the flux enters a method that determines the amount of phase being changed as well as the direction of this process. In this paper, the so-called adjacent growth method \cite{Kang2010,matthias_ehrhardt_lattice_2013} is used. It keeps track of the solid fraction $\phi$ for each cell during the simulation, where the change of $\phi$, i.e.\ $\Delta \phi$, is determined as
\begin{equation}
    \label{eq:solidFractionUpdate}
    \Delta \phi = V_\mathrm{m} a_\mathrm{m} \left( \sum_{i} g_{\bar i} - \tilde{g}_i  \right) \Delta t.
\end{equation}
Here, $V_\mathrm{m}$ is the dimensionless molar volume and $a_\mathrm{m}$ is the specific surface area. The summation term represents the reaction flux determined from the RBC scheme (cf.\ Fig.~\ref{fig:SchmaticBoundaryCondition}). In this study and as proposed by \citeA{Kang2006}, $\Delta t=1$ and $a_\mathrm{m}=1$ are used.

During the run time of the simulation $\phi$ is dynamically updated. If $\phi$ in a cell exceeds a certain threshold, the phase type of this cell is changed from solid to fluid for dissolution, or vice versa for precipitation. 
In this study, threshold values of \citeA{Kang2006} and \citeA{pedersen_improved_2014} are used. They used $\phi = 0$ for dissolution and $\phi = 1$ for precipitation. This mimics a phase change hysteresis. For precipitation, the local concentration is converted to $\phi$ and any excess is distributed to the surrounding fluid cells. The amount being distributed is weighted by the reaction rate of the cell. For dissolution, an initial concentration equal to the average concentration of the surrounding fluid cells is set and an the corresponding amount of $\phi$ is removed from the surrounding solid cells.
% OR
%For precipitation, the threshold is modified to $\phi + C\, V_\mathrm{m} = 1$. The remaining local concentration of the precipitating cell is converted to $\phi$, thus conserving mass using a compute efficient algorithm. For dissolution, the threshold is modified to $\phi - \sum_i \tilde{g}_i/\mathrm{w}_i =0$, for only the incoming $i$ populations are considered. The remaining $\phi$ of the dissolved cell is converted to the local concentration, thus avoiding large gradients while using a compute efficient algorithm.
% Used to be
% Any excess of $\phi$ during dissolution or precipitation is corrected by removing or adding, respectively, an equivalent amount of $\phi$ to the surrounding cells, weighted by the reaction rates of the cell.
%}

\section{Results} \label{sec:Results}
%{
The RBC schemes of Verhaeghe, Patel and Ju, and the new scheme are compared and verified using a variety of simulation tests. These concern the accuracy of the schemes using a robust 2D reaction-diffusion verification case with analytical solution, pattern formation during precipitation, and an encompassing case study with a reactive circular object inside a channel flow proposed by \citeA{Molins2020}. 

The different simulation tests are conducted for a broad range of advection, diffusion, and reaction regimes. They are characterized by the non-dimensional Reynolds (Re), Péclet (Pe), Damköhler (Da), and Péclet-Damköhler (PeDa) number, which are given as
\begin{equation}
\label{eq:PeDa}
\mathrm{Re} = \frac{\mathrm{U}\, \mathrm{L}}{\mathrm{\nu}}, \qquad \mathrm{Pe} = \frac{\mathrm{U}\, \mathrm{L}}{\mathrm{D}}, \qquad \mathrm{Da}=\frac{k_{\mathrm{r}}}{\mathrm{U}}, \qquad \mathrm{PeDa} = \frac{k_{\mathrm{r}} \mathrm{L}}{\mathrm{D}}.
\end{equation}
Here, $\mathrm{U}$ and $\mathrm{L}$ are the characteristic velocity and length, respectively. Note that the Péclet-Damköhler number, which opposes reaction and diffusion rates, defines the behavior of pure reaction-diffusion problems. For PeDa $\rightarrow \infty$ diffusion is limiting and for PeDa $\rightarrow 0$ it is reaction.
%}
\subsection{Analytical Reaction-Diffusion Verification} \label{sec:resultsRBCComparison}
\begin{figure}[ht!]
    \centering
    % \captionsetup{width=.6\linewidth}
    \includegraphics[width=0.55\linewidth, trim={0cm 0cm 7.5cm 0cm}, clip]{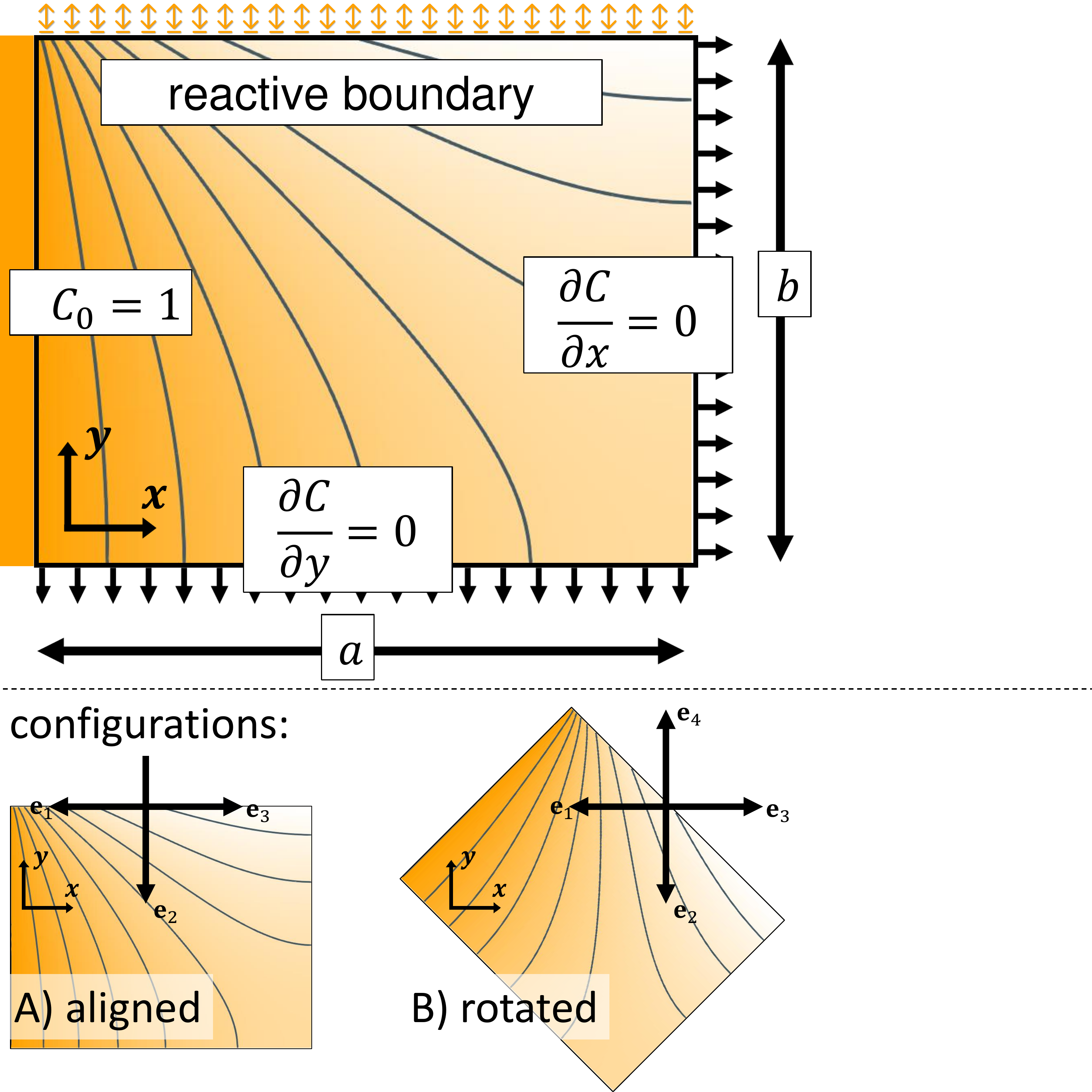}
    \caption{Simulation setup of the 2D reaction-diffusion problem. The boundary conditions are constant concentration (left), zero concentration gradient (right and bottom), and first-order equilibrium reaction (top). An exemplary concentration field and its analytical solution are indicated by the background and the contour lines, respectively. Two variants of this setup are studied; one where the boundaries are aligned with the D2Q5 grid (configuration~A) and another which is rotated clockwise by a 45$^\circ$ angle (configuration~B). }
    \label{fig:verification2DsimulationLayout}
\end{figure}

%{
A simple 2D reaction-diffusion problem is studied, for which also an analytical solution exists (cf.\ Eq.~(\ref{eq:analytical2Dverification})). The corresponding simulation setup is shown in Fig.~\ref{fig:verification2DsimulationLayout}. The bulk is solved using the BGK collision operator. The simulation considers pure diffusion of a concentration field within a rectangular domain, thus $\mathbf{u}=0$. At the left boundary, a constant concentration $C_0$ is defined. The bottom and right boundaries are set to $\nabla C=0$. At the top boundary a first-order equilibrium reaction is modelled using either the scheme of Verhaeghe, the schemes of Patel and Ju, or the new scheme.

The analytical solution to this problem \cite[p.~167]{carslaw_conduction_1986} using the formulation of \citeA{Kang2006} is
\begin{equation}
    \label{eq:analytical2Dverification}
    C^{\,(\mathrm{analyt})}(x,y)= (C_0 -C_{\mathrm{eq}}) \sum_{n=1}^\infty \left(
    \frac{ \sin{(\beta_n b)} }{N^2_n \beta_n} \frac{\cosh{(\beta_n(x-a))}}{\cosh{(\beta_n a})} \cdot \cos{(\beta_n y)} 
    \right) + C_{\mathrm{eq}},
\end{equation}
where $C^{\,(\mathrm{analyt})}(x,y)$ is the local analytical concentration, $C_{\mathrm{eq}}$ the equilibrium concentration at the top boundary, and $a$ and $b$ are the width and height, respectively. In this paper, the sum in Eq.~(\ref{eq:analytical2Dverification}) is evaluated up to n=100. The parameters $N^2_n$ is given as
\begin{equation}
    N^2_n = \frac{b}{2}\left( 1 + \frac{\sin{(2\beta_n b)}}{2\beta_n b} \right),
\end{equation}
and $\beta_n$ are the solutions to the following transcendental
\begin{equation}
    \label{eq:Analytical_PeDa}
    (\beta_n b) \tan{(\beta_n b)} = \frac{k_{\mathrm{r}} b}{\mathrm{D}} = \mathrm{PeDa}.
\end{equation}

Simulations are conducted for PeDa numbers in the range of PeDa\,=\,$[10^{-1},10^5]$ and for the two different configurations shown in Fig.~\ref{fig:verification2DsimulationLayout}: 1) Configuration~A with walls that are perfectly aligned with the grid orientation. 2) Configuration~B in which the setup is rotated by 45$^\circ$ with respect to the grid. This enables the isolated analysis of the correction factor $\gamma$ and the impact of the wall normal term $(\mathbf{e}_{\bar i}\cdot\mathbf{n})$ on the simulation result.

The accuracy of the simulations, i.e.\ the deviation of the simulation data and the analytical solution, is quantified using the mean absolute error (MAE)
\begin{equation}
   \label{eq:avgAbsError}
   \mathrm{MAE} = \sum_{x=0}^a \sum_{y=0}^b \Big| C^{\,(\mathrm{sim})}(x,y) - C^{\,(\mathrm{analyt})}(x,y) \Big| \,.
\end{equation}

If not stated otherwise, the values given in Tab.~\ref{tab:DefaultParameters} are used for the simulation. Values are given in non-dimensional units, i.e.\ lattice meter [lm], lattice second [ls], and lattice concentration [lc]. The value of $\mathrm{PeDa}$ is varied by changing the Da number only, while keeping $\tau_\mathrm{AD}$ constant in all simulations. 

\begin{table}[ht!]
\centering
\caption{\textit{Parameter set for the 2D reaction-diffusion case }}
\label{tab:DefaultParameters}
\begin{tabular}{ r l r l }
\multicolumn{2}{r}{Parameter} & Value & Unit\\ \hline 
Width             & $a$ & 100 & $\mathrm{lm}$ \\
Height            & $b$ & 80  & $\mathrm{lm}$ \\
Péclet Number     & Pe & 1    & $\mathrm{-}$ \\
Inlet Concentration & $C_0$ & 1 & $\mathrm{lc}$ \\
Equilibrium Concentration &$C_\mathrm{eq}$ & 0 & $\mathrm{lc}$ \\
Resolution        & $\mathrm{N}$ & 80  & $\mathrm{lm}$ \\
Lattice Velocity  & $\mathrm{U}$ & 0.001 & $\mathrm{lm} \, \mathrm{ls}^{\,-1}$
\end{tabular}
\end{table}
% I am not sure if makes sense to include lattice units for this segment of the paper
%}

\subsubsection{Configuration~A: Aligned System}
%{
In configuration~A, the term $\mathbf{e}_i\, \cdot\, \mathbf{n}$ in Eqs.~(\ref{eq:k_i_S_Ve})$-$(\ref{eq:k_i_S_New}) simplifies to 1 for $\mathbf{e}_2$ pointing towards the simulation domain, and 0 otherwise. Thus, the implementation of the wall normal does not affect $k_i$ (cf.\ Eqs.~(\ref{eq:k_i_S_Ve})$-$(\ref{eq:k_i_S_New})) such that the schemes of Patel and Ju, and the new scheme are identical; only the scheme of Verhaeghe differs by the factor $\gamma$.

\begin{figure}[ht!]
    \centering
    % \captionsetup{width=.8\linewidth}
    \includegraphics[width=0.85\linewidth]{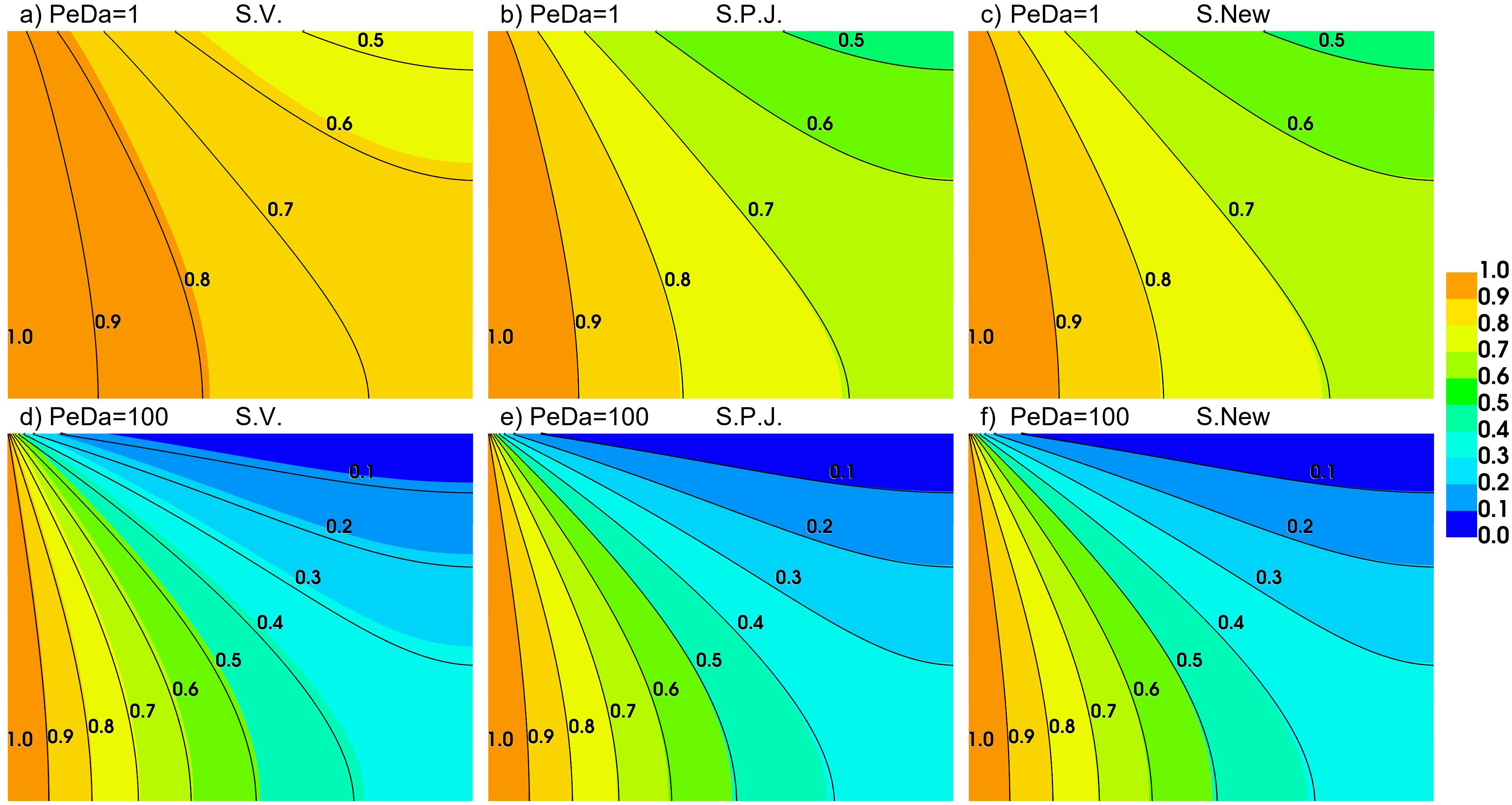}
    \caption{Simulation results of configuration~A. The concentration field is shown by the color code given in the legend. The analytical solution (cf.\ Eq.~(\ref{eq:analytical2Dverification})) is indicated by the black contour lines for the distinct values given in the plot.}
    \label{fig:verificationMergedPlots}
\end{figure}

\begin{figure}[ht!]
    \centering
    % \captionsetup{width=.8\linewidth}
    \includegraphics[width=0.85\linewidth]{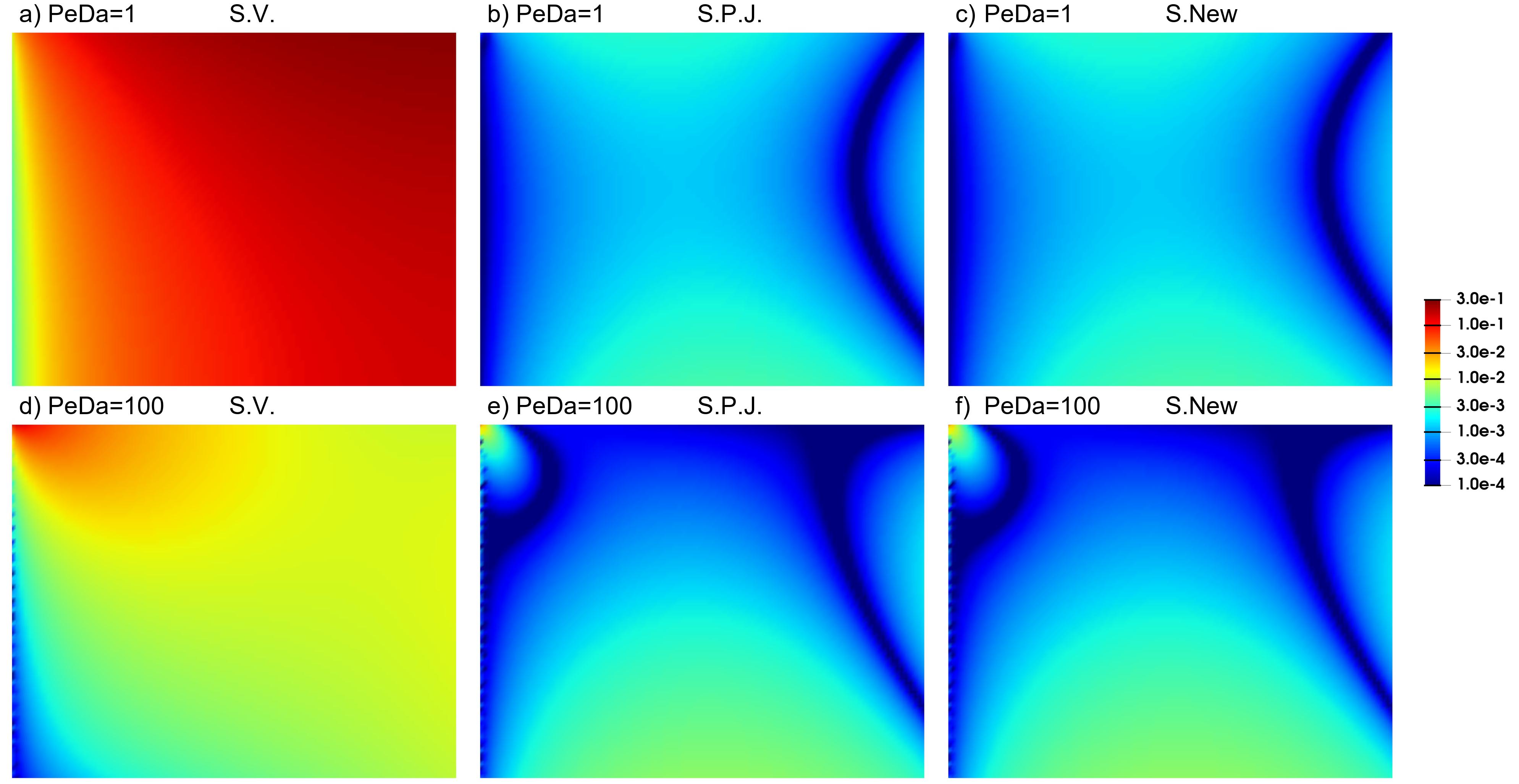}
    \caption{Absolute error of configuration~A. The error is shown by the color code given in the legend.}
    \label{fig:verificationErrorPlots}
\end{figure}

The simulation results of the concentration field are shown in Fig.~\ref{fig:verificationMergedPlots} for PeDa\,=\,1 and PeDa\,=\,100. The colors show the concentration. The analytical solution is depicted by the black contour lines for the distinct values given in the plot.
In addition, the corresponding absolute error is shown in Fig.~\ref{fig:verificationErrorPlots}. 

For the predominantly diffusion-limited case (PeDa\,=\,100), shown in Figs.~\ref{fig:verificationMergedPlots} and \ref{fig:verificationErrorPlots}~d)~to~f), all RBC schemes are at least in good agreement with the analytical solution. While for the scheme of Verhaeghe the absolute error is $<10^{-2}$ over large areas of the simulation domain, the schemes of Patel and Ju, and the new scheme are even more accurate (absolute error $<10^{-3}$). This is, however, different for the predominantly reaction-limited case (PeDa\,=\,1), shown in Figs.~\ref{fig:verificationMergedPlots} and \ref{fig:verificationErrorPlots}~a)~to~c). There, the scheme of Verhaeghe shows a significant deviation from the analytical solution, while the schemes of Patel and Ju, and the new scheme are again in almost perfect agreement with the analytical solution.

For configuration~A, the difference from the scheme of Verhaeghe compared to the schemes of Patel and Ju, and the new scheme is solely attributed to the absence of $\gamma$. As described in the literature \cite{Ju2020,Patel2016}, including $\gamma$ ensures that the macroscopic diffusivity at the wall is captured correctly. Multiplying the PeDa number in Eq.~(\ref{eq:Analytical_PeDa}) by a factor of $1/\gamma$ results in no error when using the scheme of Verhaeghe, confirming the shift in effective PeDa number simulated.

Additionally, this setup is used to confirm that integrating RBC with the rest fraction method and TRT collision operator is valid. The resultant MAE increase is less than 1\% (cf.\ SI Text~2), and thus does not result in significant additional errors. % \ref{sec:Appendix_restFraction}
%}

\subsubsection{Configuration~B: Rotated System}
%{
In configuration~B, primarily the influence of the wall normal on $k_i$ (cf.\ Eqs.~(\ref{eq:k_i_S_Ve})$-$(\ref{eq:k_i_S_New})) is tested. The simulation setup is similar to configuration~A, with the only difference that it is rotated clockwise by 45$^\circ$. This leads to the term $\mathbf{e}_i \cdot \mathbf{n}$ in Eqs.~(\ref{eq:k_i_S_Ve})$-$(\ref{eq:k_i_S_New}) being equal to $-\sqrt{2}/2$ for $\mathbf{e}_{1,\,2}$, i.e.\ pointing towards the simulation domain. Thus, none of the three schemes compared here are identical anymore. Deviations from the configuration~A predominately originate from the implementation of the wall normal.

\begin{figure}[ht!]
    \centering
    % \captionsetup{width=.8\linewidth}
    \includegraphics[width=0.85\linewidth]{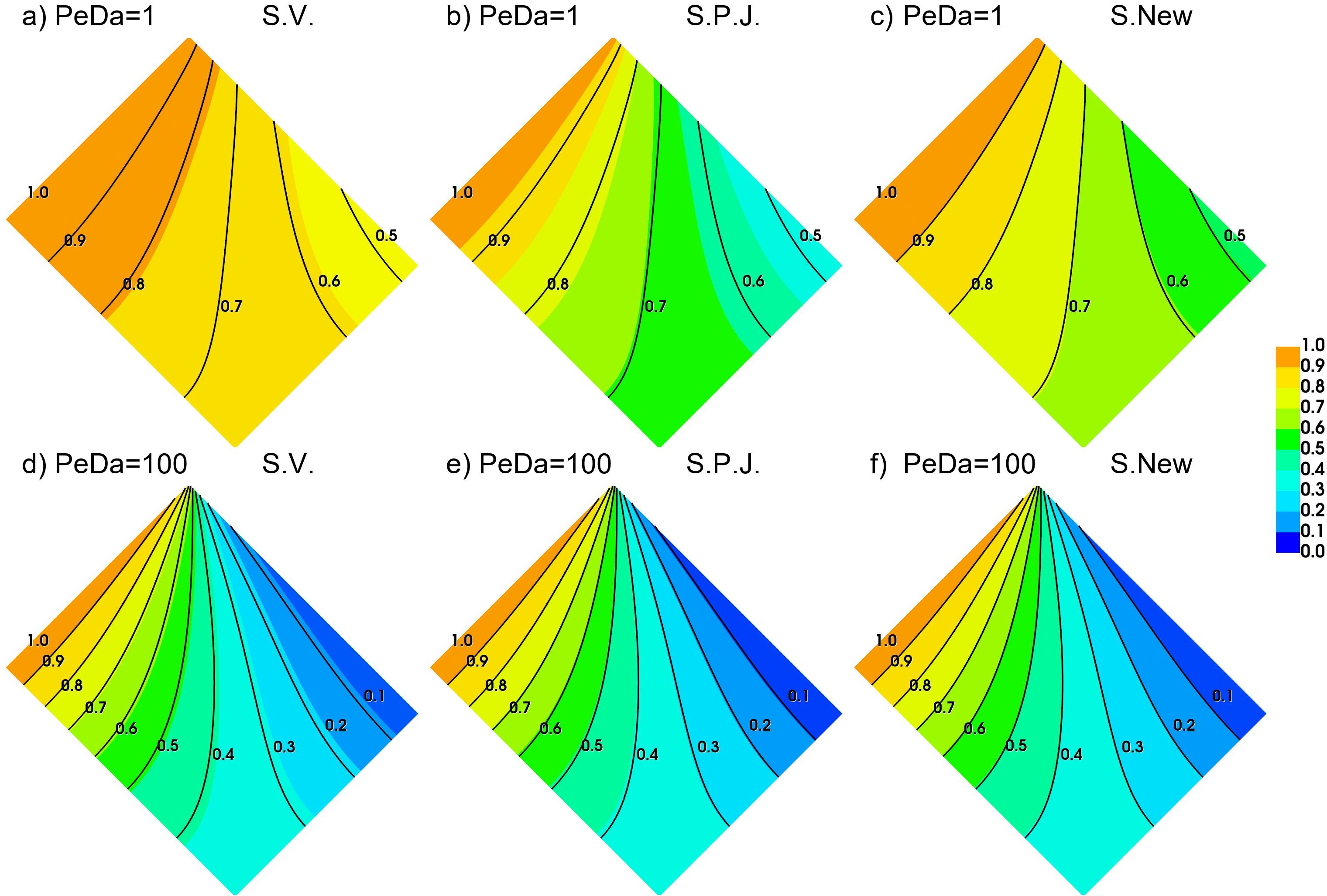}
    \caption{Simulation results of configuration~B. The concentration field is shown by the color code given in the legend. The analytical solution (cf.\ Eq.~(\ref{eq:analytical2Dverification})) is indicated by the black contour lines for the distinct values given in the plot.}
    \label{fig:verificationRotatedMergedPlots}
\end{figure}

\begin{figure}[ht!]
    \centering
    % \captionsetup{width=.8\linewidth}
    \includegraphics[width=0.85\linewidth]{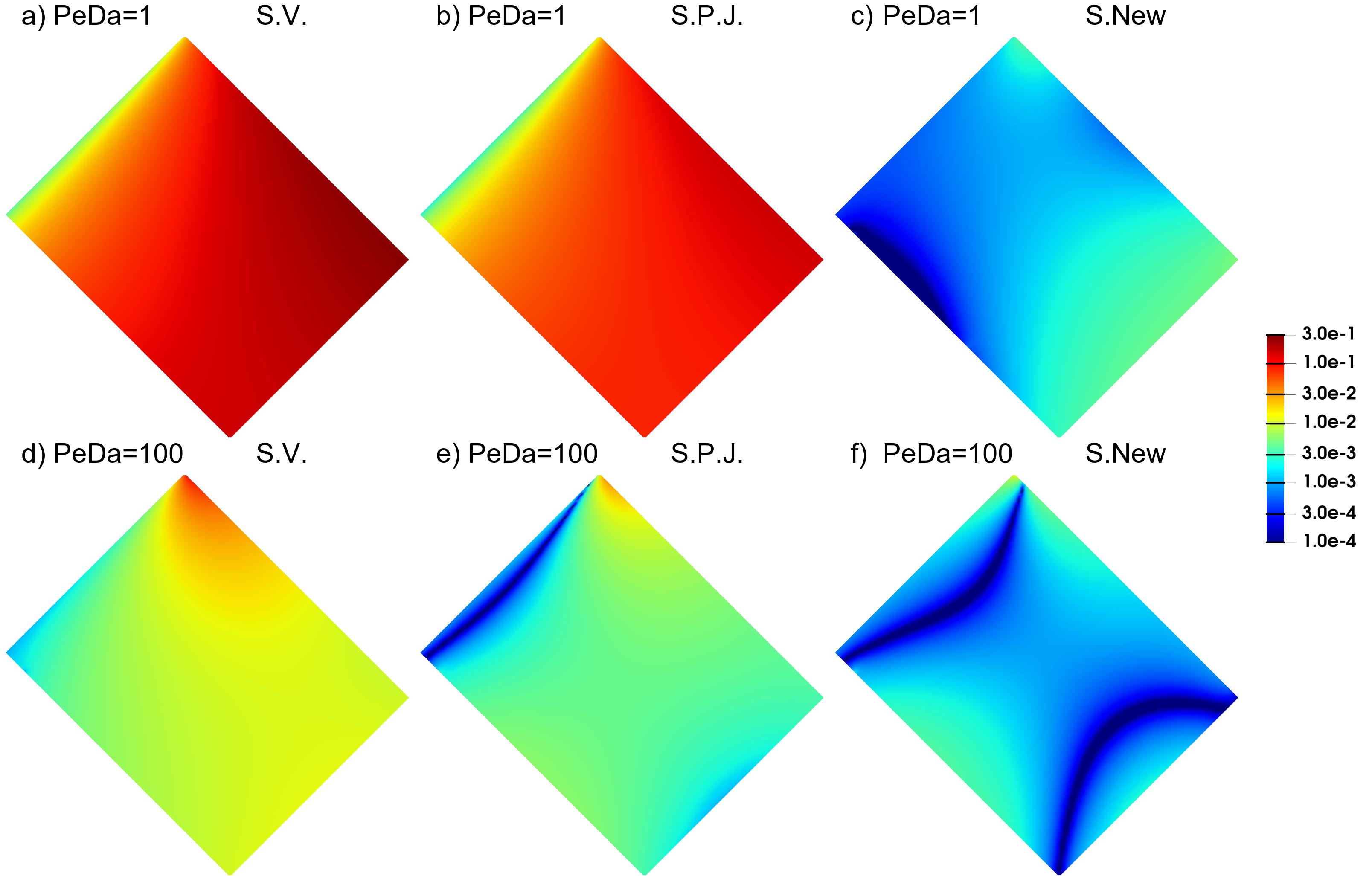}
    \caption{Absolute error of configuration~A. The error is shown by the color code given in the legend.}
    \label{fig:verificationRotatedErrorPlots}
\end{figure}

Again, for PeDa\,=\,1 and PeDa\,=\,100, the simulation results and corresponding absolute error are shown in Figs.~\ref{fig:verificationRotatedMergedPlots} and~\ref{fig:verificationRotatedErrorPlots}, respectively. The meaning of the colors and lines is identical to those from Figs.~\ref{fig:verificationMergedPlots} and~\ref{fig:verificationErrorPlots}. 

The simulation results of the scheme of Verhaeghe and the new scheme are similar to those from configuration~A. Again, multiplying the PeDa number in Eq.~(\ref{eq:Analytical_PeDa}) by a factor of $1/\gamma$, to match the analytical PeDa to the simulated PeDa, results in no error of the the scheme of Verhaeghe. However, a large deviation from the analytical solution is observed for the schemes of Patel and Ju especially at low PeDa values. 

%}

\subsubsection{Configuration~A \& B for Extended PeDa Range} \label{sec:ResultsSchemePeDaLimit}
\begin{figure}[ht!]
    \centering
    % \captionsetup{width=.8\linewidth}
    \includegraphics[width=0.9\linewidth]{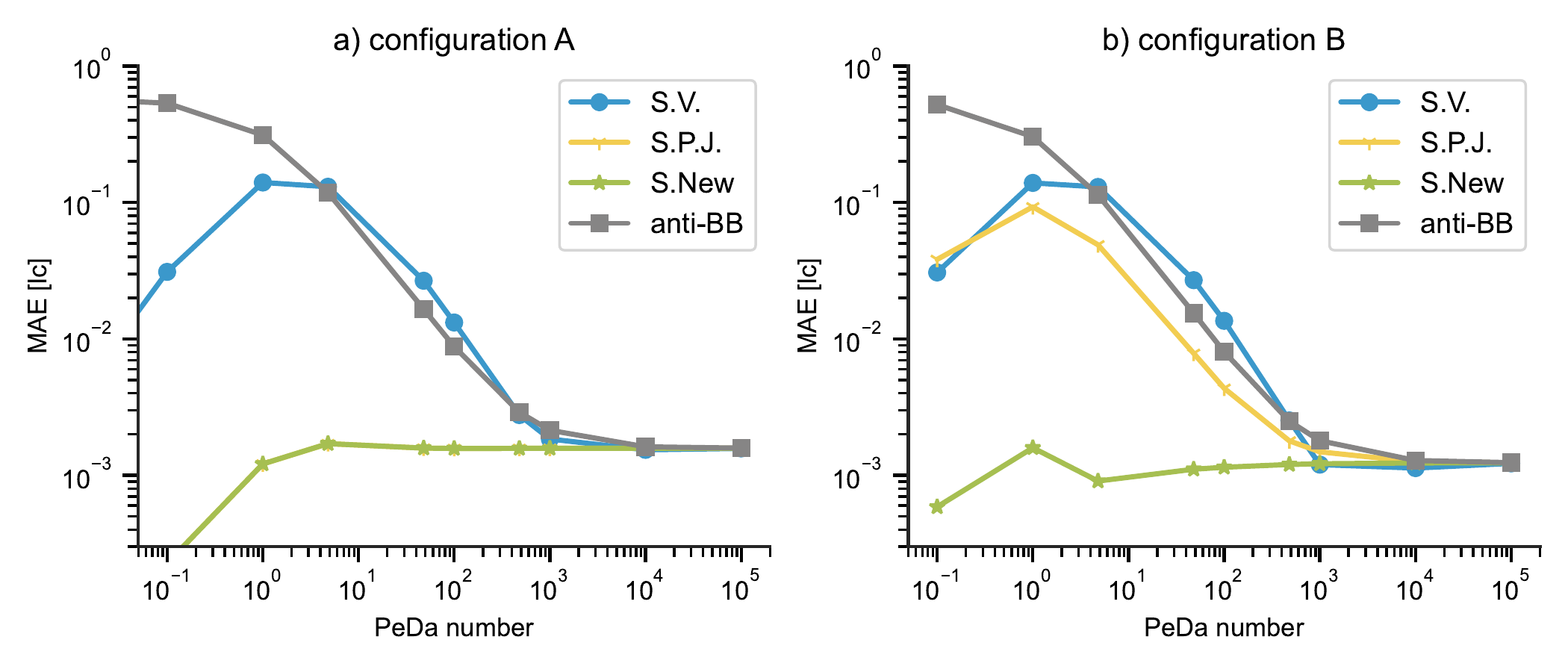}
    \caption{MAE analysis of configuration~A and B with extended PeDa range. The MAE of the schemes of Patel and Ju, and the new scheme are identical for configuration~A. }
    \label{fig:PedaToInfinity}
\end{figure}

%{
The MAE (cf.\ Eq.~(\ref{eq:avgAbsError})) of the RBC schemes are summarized in Fig.~\ref{fig:PedaToInfinity} for both configuration~A and B over a vast range of PeDa values. When  PeDa\,$\rightarrow\infty$, the RBC schemes approach the anti-BB method. This was shown in the literature \cite{Verhaeghe2006,Ju2020} and is further elaborated in the SI Text~S1.3. % \ref{sec:Appendix_ResultsSchemePeDaLimit}
Note, that the anti-BB sets the equilibrium concentration at the boundary. Therefore, this approximation is expected to have significant errors at low PeDa values. Still, it is included as reference in the extended MAE plots.

For configuration~A (cf.\ Fig.~\ref{fig:PedaToInfinity}~a)), it clearly shows how the scheme of Verhaeghe has significant errors for $ 10^{-1} \lesssim$ PeDa $\lesssim 10^2$, with an decreasing error for when PeDa$\rightarrow\infty$. On the other hand, the schemes of Patel and Ju, and the new scheme are barely affected by the choice of PeDa. This highlights the impact of $\gamma$. The MAE decreases for all schemes in the reaction-limited case, i.e\ PeDa$\;<1$, which has the trivial solution of $C(x,y)=C_0$, with the RBC approaching bounce-back behavior.

For configuration~B (cf.\ Fig.~\ref{fig:PedaToInfinity}~b)), the scheme of Verhaeghe and the new scheme behave the same as in configuration~A (cf.\ Fig.~\ref{fig:PedaToInfinity}~a)). However, the scheme of Patel and Ju shows a significant error, due to their wall-normal implementation. In contrast, the new scheme developed in the current study, is almost not affected by the choice of PeDa for both configuration~A and B. 

The MAE values for RBC and anti-BB schemes converge for PeDa$\;>10^3$ in configuration~A and for PeDa$\;>10^4$ in configuration~B. Thus, simulations with PeDa$\;>10^4$ show similar results for all aforementioned RBC schemes, i.e.\ anti-BB, the schemes of Verhaeghe, Patel and Ju, and the new scheme.

%}

\subsection{Pattern Formation During Precipitation} \label{sec:resultsPrecipitation}
%{
This test case is motivated by the work of \citeA{pedersen_improved_2014}. 
They investigated pattern formation of grain growth during precipitation and its dependence on the grid orientation. For their study they used the scheme of Verhaeghe.

The simulation domain is circular with a diameter of 300 grid cells. This is realized by adding boundary cells along this circle in a square domain of size 300 lm $\times$ 300 lm cells. On the boundary the concentration is constant, i.e.\ $C_{\mathrm{circ}}=1$. An initial seed is placed in the center of the domain. Its shape is defined by the parametric equation
\begin{equation}
    \begin{array}{c}{
    {x(s)=[0.1+0.02\,\mathrm{cos}(8\pi s)]\,\mathrm{cos}(2\pi s),}}\\ 
    {{y(s)=[0.1+0.02\,\mathrm{cos}(8\pi s)]\,\mathrm{sin}(2\pi s),}}
    \end{array}
\end{equation}
with $s$ as the parameter. Two variants of this setup are simulated: One where the initial seed is aligned with the grid orientation, and another where the initial seed is rotated clockwise by 19$^\circ$. The boundaries of the seed are defined by a RBC with $C_{\mathrm{eq}}=0$ and $k_{\mathrm{r}}=0.0014$ (cf.\ Eq.~(\ref{eq:general_RBC})).  The bulk is solved using the BGK collision operator. Pure diffusion is considered such that $\mathbf{u}=0$ within the domain. To compute the wall normal, the simple and efficient isotropic finite difference method \cite{kumar_isotropic_2004} is used. 

\begin{figure}[ht!]
    \centering
    % \captionsetup{width=.8\linewidth}
    \includegraphics[width=0.9\linewidth]{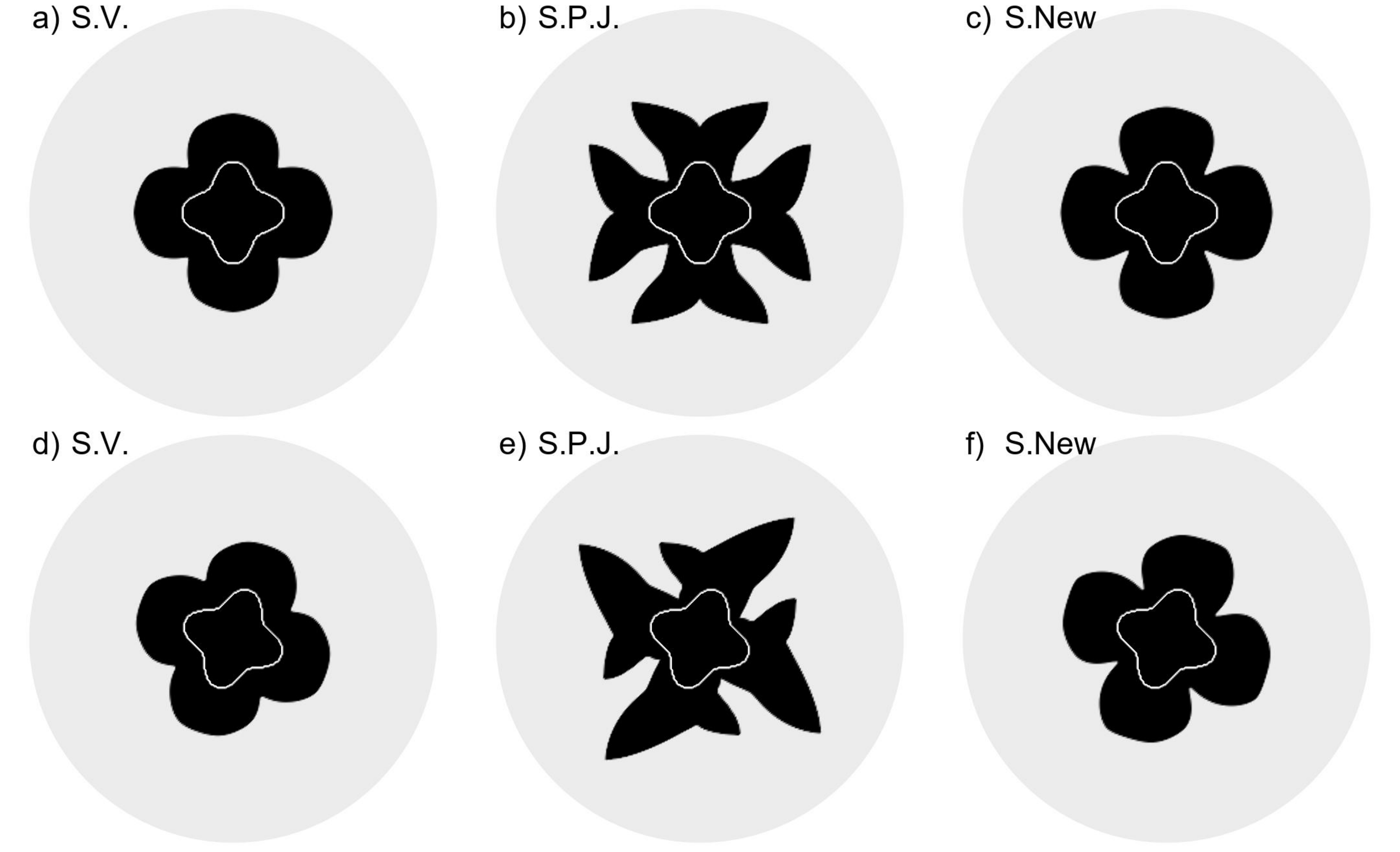}
    \caption{Pattern formation during precipitation. The final grains are shown in black. The initial grain seeds and the simulation domain are indicated by the white line and the gray-shaded background, respectively.}
    \label{fig:grainGrowth}
\end{figure}

The results determined with the scheme of Verhaeghe, the schemes of Patel and Ju, and the new scheme are shown in Fig.~\ref{fig:grainGrowth}. The resulting patterns when using the scheme of Verhaeghe and the new scheme are similar, as shown in Fig.~\ref{fig:grainGrowth}~a)~\&~d) and Fig.~\ref{fig:grainGrowth}~c)~\&~f), respectively. However, for the scheme of Verhaeghe the growth is less pronounced which is due to the effective PeDa shift. A correction of this shift further improves the accordance of both schemes. In contrast, the pattern resulting from the schemes of Patel and Ju shows dendritic behavior and grid dependence, shown in Fig.~\ref{fig:grainGrowth}~b)~\&~e). The reason is due to the implementation of the wall normal in $k_{i}^{(\,\mathrm{S.P.J.})}$ (cf.\ Eq.~(\ref{eq:k_i_S2})), leading to faster reactions in diagonal walls compared to straight walls. 

%}

\subsection{Reaction in a Channel Flow} \label{sec:resultsMolins}
%{
This case study considers a reactive object in a channel flow, i.e.\ under the influence of advection. It is motivated by the extensive comparison paper by \citeA{Molins2020} in which they compared five different simulation methods (Chombo-Crunch, dissolFoam, LBM, OpenFoam, and Vortex). Both \citeA{Molins2020} and \citeA{Ju2020} used the same dissolution experiment of \citeA{soulaine_mineral_2017} to validate their methods.

\begin{figure}[!ht]
    \centering
    % \captionsetup{width=.8\linewidth}
    \includegraphics[width=0.85\linewidth, trim={0cm 8.5cm 0cm 6.5cm}, clip]{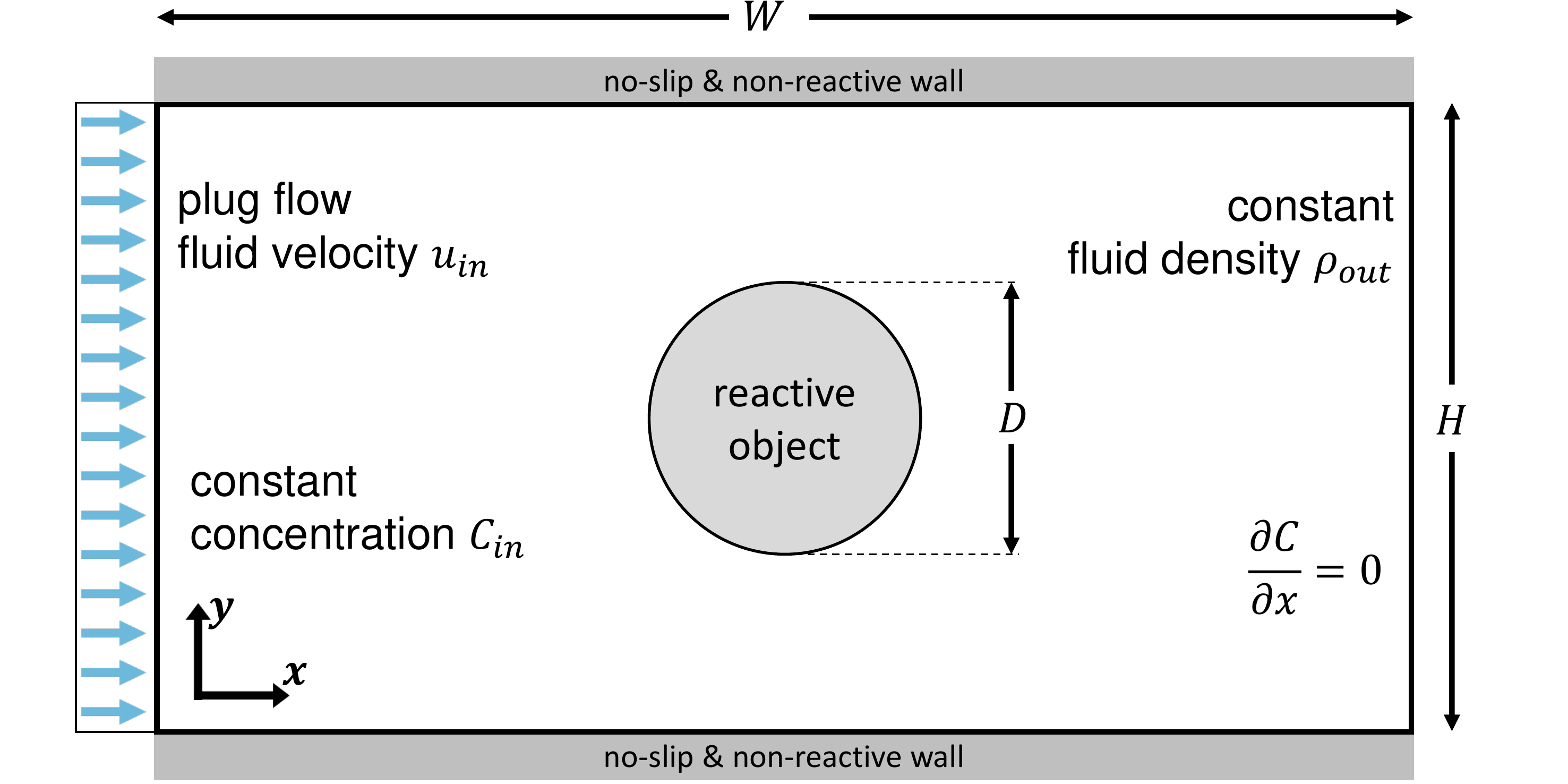}
    \caption{Simulation setup of the 2D reactive object in channel flow from \protect\citeA{Molins2020}. The boundary conditions for 1) the fluid lattice are constant plug flow (left), constant density (right) and bounce-back walls (top, bottom, object), and for 2) the scalar lattice are constant concentration (left), zero concentration gradient (right), bounce-back walls (top, bottom) and RBC (object).}
    \label{fig:molinsSetup}
\end{figure}

The simulation setup is shown in Fig.~\ref{fig:molinsSetup}. It is 2D and consists of a channel with height $H=0.05$\,cm and width $W=0.1$\,cm, simulated with resolution $600\,\mathrm{lm}\times\,1200\,\mathrm{lm}$ and a lattice velocity of $\mathrm{U}_{LB} = 0.001$ lm ls$^{-1}$. The channel is initially filled with a fluid at rest and with constant concentration equal to that of the inlet. A circular reactive object with diameter $D=0.02$\,cm is placed in the middle of the channel. The boundary conditions for the carrier fluid are 1) bounce-back walls at the top, bottom, and at the object's surface, 2) plug velocity profile $u_{in}=0.12$\,cm/s at the inlet on the left, and 3) fixed density $\rho_{out}=1$ at the outlet on the right. The boundary conditions for the concentration field are 1) bounce-back walls at the top and bottom, 2) constant concentration at the inlet $C_{in} = 10$\,mol/m$^3$, 3) zero concentration gradient at the outlet, and 4) a RBC at the object's surface. The RBC is varied between the three schemes and the anti-BB. Large ratios of Pe to Re are realized using the TRT collision operator and rest fraction method for the concentration lattice. The following values are chosen: $J_0 = 0.99$ for Pe=600 and $J_0=1/3$ for Pe=6. First, the flow of the carrier fluid, i.e.\ the NS lattice, was simulated until steady state was achieved. Then, the concentration field, i.e.\ the AD lattice, was solved and the average reaction rate of the object was determined as
\begin{equation}
   \label{eq:avgReactionRateMolins}
   R_{\mathrm{avg}} =  \frac{1}{\pi D} \left( \int_{\mathrm{outlet}} C u_x \cdot dy - C_{in}u_{in}H   \right).
\end{equation}
Here the integral is performed on the outlet.

Simulations were conducted for Re\,=\,0.6 and different combinations of Pe\,=\,[6, 600] and Da\,=\,[0.178, 17800]. These are: case 1) Pe = 600, Da = 178 (diffusion limited); case 2) Pe = 600, Da = 17800 (diffusion limited); case 3) Pe = 6, Da = 178 (advection-diffusion limited); and case 4) Pe = 6, Da = 0.178 (advection-reaction-diffusion limited). 
As is shown in the following, the cases 1) to 3) are not suited to differentiate the schemes, since they are not reaction limited.

\begin{figure}[ht!]
    \centering
    % \captionsetup{width=.8\linewidth}
    \includegraphics[width=0.85\linewidth]{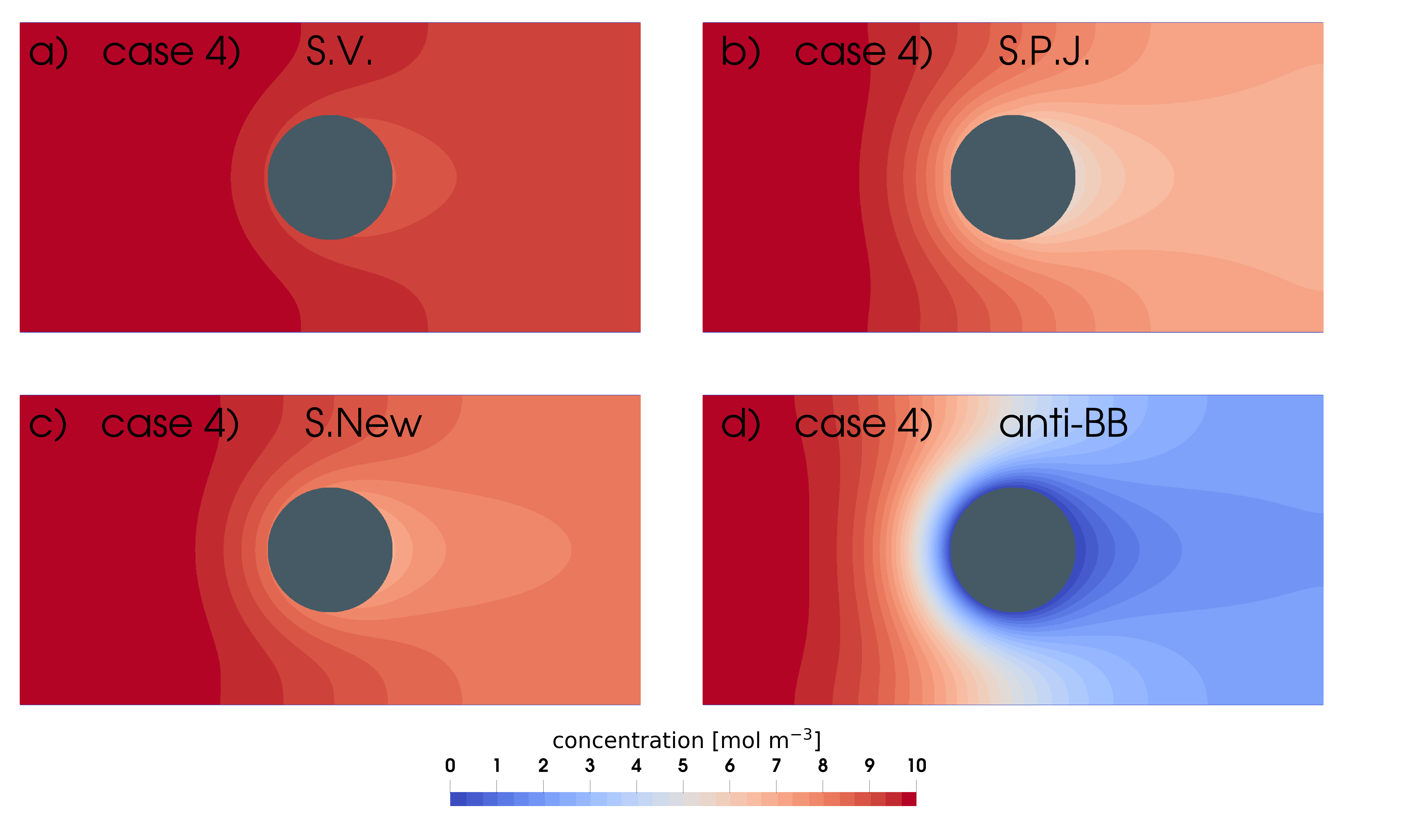}
    \caption{Concentration field in the vicinity of a reactive object (grey circle) in a channel flow simulated for case 4). Using a) the scheme of Verhaeghe, b) the schemes of Patel and Ju, c) the new scheme, and d) the anti-BB scheme, which sets the concentration to the equilibrium concentration. The concentration is shown by the color code given in the legend. }
    \label{fig:molinsResultsCase4}
\end{figure}

A qualitatively comparison of the results of case 4) for the schemes is given in Fig.~\ref{fig:molinsResultsCase4}. A comparison between the cases 1) to 4) for the new scheme is shown in Fig.~1 in the SI.
In Case~4) when using the scheme of Verhaeghe (cf.\ Fig.~\ref{fig:molinsResultsCase4}~a)), the simulation is strongly reaction limited, as indicated by the high concentration around the object's surface. In contrast, the anti-BB scheme (cf.\ Fig.~\ref{fig:molinsResultsCase4}~d)) shows the case where reaction is not limiting at all. Both the new scheme and the schemes of Patel and Ju (cf.\ Fig.~\ref{fig:molinsResultsCase4}~b)\,\&\,c), respectively)
fall between the two extremes.

The numerical results of $R_{\mathrm{avg}}$ are given in Table~\ref{tab:MolinsVerification} together with the values from \citeA{Molins2020}. There, ranges are given as different methods were used to determine $R_{\mathrm{avg}}$ (cf.\ SI Text~S3). % \ref{sec:Appendix_reactiveObject}
Overall, the RBC schemes agree well with the results of Molins {et al.}, where available, indicating good integration of the RBC schemes with the rest fraction method. 

\begin{table}[ht!]
\centering
\caption{\textit{Results of $R_{\mathrm{avg}}$ of a Reactive Object in a Channel Flow}}
\label{tab:MolinsVerification}
\begin{tabular}{c r l | l l l l l }
\multicolumn{3}{c}{Input} &  \multicolumn{5}{c}{$R_{\mathrm{avg}}$ $[10^{-8}\; \mathrm{mol/(cm^2\cdot s)}]$} \\
Case & Pe & Da & Molins~{et al.} & S$_{\mathrm{Ve}}$ & S$_{\mathrm{P.J.}}$ & S$_{\mathrm{New}}$ & Anti-BB \\ \hline
1) & 600 & 178   & [4.18, 4.57] & $4.48$ & $4.66$ & $4.64$ & $4.71$ \\
2) & 600 & 17800 & [3.88, 4.79] & $4.70$ & $4.70$ & $4.70$ & $4.71$ \\ 
3) & 6   & 178   & [58.6, 90.5] & $75.5$ & $75.9$ & $75.8$ & $76.0$ \\
4) & 6   & 0.178 & N/A          & $8.05$ & $29.1$ & $18.5$ & $76.0$ \\
\end{tabular}
\end{table}
Table~\ref{tab:MolinsVerification} shows that for the cases 1) to 3), the simulations are not reaction limited, resulting in the RBCs behaving similar to the anti-BB. Thus, cases in which the PeDa $\geq 1000$ are not suited for validation of RBC schemes. The experiment of \citeA{soulaine_mineral_2017} has a PeDa\,=\,3000.
In case 4), the reaction is not limiting anymore, as indicated by the much higher anti-BB result. Here, the result of the various RBC schemes diverge. Compared to the new scheme, the scheme of Verhaeghe underestimates and the schemes of Patel and Ju overestimates $R_{\mathrm{avg}}$ by the significant factors of 0.43 and 1.6 respectively. 

%}

\section{Conclusion} \label{sec:Conclusion}
%{
A new general local reactive boundary condition (RBC) scheme was presented that accurately captures first-order equilibrium reactions for a wide range of Péclet (Pe) and Damköhler (Da) numbers for complex geometries. It combines aspects of previous RBCs from \citeA{Verhaeghe2006}, \citeA{Ju2020}, and \citeA{Patel2016}, but overcomes their deficiencies. This means, the new RBC scheme does not suffer from a $\tau$-dependent wall diffusion and accurately considers the wall normal for complex geometries and surfaces that are not aligned with the simulation grid.

In this study, all aforementioned RBC schemes were first reformulated to a similar notation. Then, all RBC schemes were tested and compared using three different verification cases. These were chosen such that they were relevant, representative, and covered important features of RBCs and their applications: 
1) A robust 2D reaction setup purely driven by diffusion. The problem has an analytical solution and the RBC schemes were studied over a wide range of PeDa values as well as under grid rotation. 
2) Pattern formation in a precipitation process. Here, especially the impact of rotational variance and the behavior of the wall normal was studied. 
3) An advanced reaction setup including advection around a reactive object in a channel flow. The RBC schemes were coupled to the rest fraction method, a wide range of PeDa values was studied, and the results were compared to results from the literature.

The test cases demonstrate a broad applicability and high accuracy of the new RBC scheme. For the 2D reaction setup, the new RBC scheme shows the best accordance with the analytical solutions. Additionally, it was shown that it correctly simulates wall diffusion and is invariant with respect to relaxation time and grid orientation. For the simple precipitation case, the new RBC scheme showed physically sound pattern formations. Moreover, in contrast to the other RBC schemes, it was not affected by grid rotation. For the advection case, large ratios of Péclet and Reynolds number ($\leq 1000$) were successfully simulated. Reaction rates determined using the new RBC scheme are in general accordance with results from the literature \cite{Molins2020}.

All in all, the new RBC accurately describes first-order reactions and can be applied to simulate precipitation and dissolution phenomena even for complex geometries. This is a strong advantage over other RBCs previously described in the literature. In addition, as a general and local scheme, it is easy to implement for both 2D and 3D simulations, it is computationally efficient and facilitates parallel computation. The new RBC scheme is missing some advanced features, e.g.\ moving walls, or interpolated sub-grid wall locations. However, the presented general reformulation allows those already developed features to be incorporated easily.

The new RBC can be used to study reaction processes in a broad range of research fields. Potential applications might be reactive flows through porous media, with and without dissolution and precipitation \cite{pereira_numerical_2022, zhang_porescale_2019, zhang_pore-scale_2021, zhang_influence_2021, xu_lattice_2018, jiang_pore-scale_2021}, pore structure evolution in cement manufacturing \cite{Patel2021b}, or morphological changes due to the conversion of active material in energy storage systems \cite{fang_pore-scale_2021}. Especially the last topic will be in the focus of our future work. 

%}

%%%%%%%%%%%%%%%%%%%%%%%%%%%%%%%%%%%%%%%%%%%%%%%
%Authors should include an Availability Statement for the software that has a significant impact on the research. Details and templates are in the Availability Statement section of the Data and Software for Authors Guidance: \url{https://www.agu.org/Publish-with-AGU/Publish/Author-Resources/Data-and-Software-for-Authors#availability}

\section*{Open Research} \label{sec:OpenResearch}

An extended version of the \textbf{Pa}rallel \textbf{La}ttice \textbf{Bo}ltzmann \textbf{S}olver (version 2.3.0) --- short Palabos --- was used for all LBM simulations in this study. The original version of Palabos is preserved at \url{https://palabos.unige.ch/}, available via GNU Affero General Public License version 3 without login required and developed openly at \url{https://gitlab.com/unigespc/palabos} \cite{latt_palabos_2021}. The reactive boundary conditions and the rest fraction method were implemented separately.

\section*{Acknowledgements}
The authors gratefully acknowledge financial support by the Federal Ministry of Education and Research (BMBF) under Grant No.03XP0491A. M.L. and B.K. gratefully acknowledge financial support from the European Union's Horizon 2020 Research and Innovation Programme within the project ``DEFACTO'' under the grant number 875247. The simulations were carried out on the Hawk at the High Performance Computing Center Stuttgart (HLRS) under the grant LaBoRESys, and on bwHPC JUSTUS 2 at the University Ulm under the grant INST 40/467-1 FUGG.

%%%%%%%%%%%%%%%%%%%%%%%%%%%%%%%%%%%%%%%%%%%%%%%
%%%                 APPENDIX                %%%
%%%%%%%%%%%%%%%%%%%%%%%%%%%%%%%%%%%%%%%%%%%%%%%

\begin{appendices}
\section{LBM Velocity Sets} \label{sec:App_VelocitySet}
%{
This paper discusses the 2D implementation of the RBC only, as the extensions to 3D are simple and straightforward. 
The velocity sets used for the 2D implementations are D2Q9 and D2Q5 and are given in the following:

\noindent\textbf{D2Q9}
\begin{eqnarray}
    \label{eq:D2Q9}
    &\{\mathbf{e}_i\}&={\left\{\mathbf{e}_{0}, \mathbf{e}_{1}, \mathbf{e}_{2}, \mathbf{e}_{3}, \mathbf{e}_{4}, \mathbf{e}_{5}, \mathbf{e}_{6}, \mathbf{e}_{7}, \mathbf{e}_{8}\right\}} \nonumber \\
    &&=\frac{\Delta x}{\Delta t}\left\{\begin{array}{rrrrrrrrr}
        0 & -1 & -1 & -1 &  0 &  1 & 1 & 1 & 0 \\
        0 &  1 &  0 & -1 & -1 & -1 & 0 & 1 & 1
    \end{array}\right\} \\
    &\{\mathrm{w}_i\}& = {\left\{ \frac{4}{9}, \frac{1}{36}, \frac{1}{9}, \frac{1}{36}, \frac{1}{9}, \frac{1}{36}, \frac{1}{9}, \frac{1}{36}, \frac{1}{9}  \right\}}   \nonumber
\end{eqnarray}

\noindent\textbf{D2Q5}
\begin{eqnarray}
    \label{eq:D2Q5}
    &\{\mathbf{e}_i\}&={\left\{\mathbf{e}_{0}, \mathbf{e}_{1}, \mathbf{e}_{2}, \mathbf{e}_{3}, \mathbf{e}_{4}\right\}}\nonumber \\
    &&=\frac{\Delta x}{\Delta t}\left\{\begin{array}{rrrrr}
        0 & -1 &  0 & 1 & 0 \\
        0 &  0 & -1 & 0 & 1
    \end{array}\right\} \\
    &\{\mathrm{w}_i\}& = {\left\{ \frac{1}{3}, \frac{1}{6}, \frac{1}{6}, \frac{1}{6}, \frac{1}{6} \right\}}   \nonumber
\end{eqnarray}
In both Eqs.~(\ref{eq:D2Q9})~\&~(\ref{eq:D2Q5}) the speed of sound is $c_{\mathrm{s}}=1/\sqrt{3}\ \ \Delta \mathbf{x}/\Delta t$. When using the rest fraction method, both weights and speed of sound change (cf.\ Section~\ref{sec:RestFraction}).
%}
\end{appendices}

%% ------------------------------------------------------------------------ %%
%% References and Citations

\bibliography{ms}

\end{document}

%% ------------------------------------------------------------------------ %%
%
%  TITLE
%
%% ------------------------------------------------------------------------ %%

\begin{center}
     Supporting Information for\\[1em]
    \textbf{ General Local Reactive Boundary Condition for Dissolution and Precipitation \\ Using the Lattice Boltzmann Method}
\end{center}

%% ------------------------------------------------------------------------ %%
%
%  AUTHORS AND AFFILIATIONS
%
%% ------------------------------------------------------------------------ %%

\vspace{1em}
\begin{center}
     J. Weinmiller\affil{1,2},
     M. P. Lautenschlaeger\affil{1,2},
     B. Kellers\affil{1,2}, \\ 
     T. Danner\affil{1,2},
     A. Latz\affil{1,2,3}
\end{center}

\affiliation{1}{German Aerospace Center (DLR), Institute of Engineering Thermodynamics, 89081 Ulm, Germany}
\affiliation{2}{Helmholtz Institute Ulm for Electrochemical Energy Storage (HIU), 89081 Ulm, Germany}
\affiliation{3}{Ulm University (UUlm), Institute of Electrochemistry, 89081 Ulm, Germany}

\vspace{1em}
\begin{center}
    \small Corresponding author: J. Weinmiller (julius.weinmiller@dlr.de)
\end{center}
\vspace{3em}

%% ------------------------------------------------------------------------ %%
%
%  TEXT
%
%% ------------------------------------------------------------------------ %%

\section*{Contents of this file}
\begin{itemize}
    \item Text S1 to S3
    \item \vspace{-0.5em} Figure S1
\end{itemize}

\section*{Introduction}
This supporting information provides details on the reactive boundary conditions schemes, their reformulation and behavior in the limit. Additionally, the supporting information contains a section on the aggregation of \citeA{Molins2020} simulation data.

\clearpage

\section{Reactive Boundary Condition Expressions}
\label{sec:Appendix_BoundariesExpression}

In the following, the derivation of the reactive boundary condition (RBC) of \citeA{Verhaeghe2006} is given in detail. In addition, the approach to reformulate the RBC schemes of \citeA{Ju2020} and \citeA{Patel2016} to the notation of Verhaeghe et al. is shown. As in the main text, only the first authors are mentioned in future discussions. The nomenclature and symbols are chosen in a consistent manner with respect to the current paper.

\subsection{Derivation of the RBC scheme of Verhaeghe}
\label{sec:Appendix_derivationVerhaegheBoundary}

The scheme of Verhaeghe is based on a Dirichlet boundary condition implemented using the anti bounce-back method (cf.\, Eq.~(\ref{eq:anti_BB})), a wall flux analogous to moving walls (cf.\, Eq.~(\ref{eq:fluxBC})) \cite{bouzidi_momentum_2001}, and a definition of the reactive flux (cf.\, Eq.~(\ref{eq:fluxFirstOrder})). The corresponding equations read:
\begin{equation}
    \label{eq:anti_BB}
    g_i + g_{\bar{i}} = 2\mathrm{w}_i C_{\mathrm{w}},
\end{equation}
with the outgoing population $g_i$, the incoming population $ g_{\bar{i}}$ and the wall concentration $C_{\mathrm{w}}$, and
\begin{equation}
    \label{eq:fluxBC}
    g_i = g_{\bar{i}} - 2\mathrm{w}_i J_{\mathrm{R}} \frac{\mathbf{c}_i \cdot \mathbf{n}}{c_{\mathrm{s}}^2},
\end{equation}
where momenta flux due to the wall movement is interpreted as concentration flux along the wall normal $J_{\mathrm{R}} \mathbf{n}_\mathrm{w}$. 
\begin{equation}
    \label{eq:fluxFirstOrder}
    J_{\mathrm{R}} = k_{\mathrm{r}} \left( C_{\mathrm{eq}} - C_{\mathrm{w}} \right),
\end{equation}
where $J_{\mathrm{R}}$ is the reactive flux defined by a first-order equilibrium reaction.
From the aforementioned equations, the scheme of Verhaeghe is derived as given in the following.
Solving Eq.~(\ref{eq:anti_BB}) for $C_{\mathrm{w}}$ and inserting it into Eq.~(\ref{eq:fluxFirstOrder}) gives
\begin{equation}
    \label{eq:fluxWall}
    J_{\mathrm{R}} = k_{\mathrm{r}} \left( C_{\mathrm{eq}} - \frac{g_i + g_{\bar{i}}}{ 2\mathrm{w}_i} \right).
\end{equation}

Inserting Eq.~(\ref{eq:fluxWall}) into Eq.~(\ref{eq:fluxBC}) results in
\begin{equation}
    \label{eq:fluxBCPreSolve}
    g_i = g_{\bar{i}} - 2\mathrm{w}_i k_{\mathrm{r}} \left( C_{\mathrm{eq}} - \frac{g_i + g_{\bar{i}}}{ 2\mathrm{w}_i} \right) \frac{\mathbf{c}_i \cdot \mathbf{n}}{c_{\mathrm{s}}^2},
\end{equation}
or
\begin{equation}
    \label{eq:fluxBCMidSolve1}
    g_i = g_{\bar{i}} - 2\mathrm{w}_i k_i \left( C_{\mathrm{eq}} - \frac{g_i + g_{\bar{i}}}{ 2\mathrm{w}_i} \right),
\end{equation}
when applying $k_{i}^{(\mathrm{V})}  = c_{\mathrm{s}}^{-2} k_{\mathrm{r}}\ \mathbf{c}_i \cdot \mathbf{n}$.

Solving Eq.~(\ref{eq:fluxBCMidSolve1}) for $g_{\bar{i}}$ leads to 
\begin{equation}
    \label{eq:fluxBCSolveWrongSign}
    g_{\bar{i}}  =  -\frac{k_i}{1 - k_i} 2 \mathrm{w}_i C_{\mathrm{eq}} +  \frac{ 1 + k_i}{1 - k_i} g_i.
\end{equation}
Finally, applying $\mathbf{c}_i = - \mathbf{c}_{\bar{i}}$ in the definition of $k_i$ results in $k_{\bar{i}}=-k_i$, and finally leads to
\begin{equation}
    \label{eq:fluxBCSolve}
    g_{\bar{i}}  =  \frac{k_{\bar{i}}}{1 + k_{\bar{i}}} 2 \mathrm{w}_i C_{\mathrm{eq}} +  \frac{ 1 - k_{\bar{i}}}{1 + k_{\bar{i}}} g_i.
\end{equation}
This final equation (cf.\, Eq.~(\ref{eq:fluxBCSolve})) is similar to the result from \citeA{pedersen_improved_2014}. 
%}

% \vspace{2em}\noindent\textbf{Text S1.2. Reformulation of the RBC of Ju and Patel} 
\subsection{Reformulation of the RBC schemes of Ju and Patel}
\label{sec:Appendix_PatelJuDeriv}
%{
Starting from the original formulations of the schemes of Ju and Patel it is shown under which conditions they can be reformulated similar to the scheme of Verhaeghe (cf.\ Eq.~(\ref{eq:fluxBCSolve})).

The original formulation of \citeA[p.69]{Patel2016} is
\begin{equation}
\label{eq:PatelBCOG}
    g_{\bar{i}} = \frac{ {k_{\mathrm{r}}C_{\mathrm{eq}}-\tilde{g}_{i}\left[{\frac{V}{A_{\mathrm{s}}}{\frac{k_{\mathrm{r}}}{2\mathrm{w}_{i}}}-\left(1-{\frac{\Delta t}{2\tau_{\mathrm{AD}}}}\right)\mathbf{e}_{\bar{i}}\cdot\mathbf{n}}\right]}} {{ \frac{1}{A_{\mathrm{s}}}{\frac{k_{\mathrm{r}}}{2\mathrm{w}_{i}}}-\left(1-{\frac{\Delta t}{2\tau_{\mathrm{AD}}}}\right)\mathbf{e}_{i}\cdot\mathbf{n}}}.
\end{equation}
Here, $V$ and $A_{\mathrm{s}}$ are the volume and surface area, respectively, of the solid at the boundary. They are unity when non-dimensionalized, resulting in
\begin{equation}
\label{eq:PatelBC}
    g_{\bar{i}} = \frac{ {k_{\mathrm{r}}C_{\mathrm{eq}}-\tilde{g}_{i}\left[{{\frac{k_{\mathrm{r}}}{2\mathrm{w}_{i}}}-\left(1-{\frac{1}{2\tau_{\mathrm{AD}}}}\right)\mathbf{e}_{\bar{i}}\cdot\mathbf{n}}\right]}} {{{\frac{k_{\mathrm{r}}}{2\mathrm{w}_{i}}}-\left(1-{\frac{1}{2\tau_{\mathrm{AD}}}}\right)\mathbf{e}_{i}\cdot\mathbf{n}}}.
\end{equation}

The original formulation of \citeA[Eq.~(33)]{Ju2020} without wall velocity for reduced lattices is
\begin{equation}
\label{eq:GuoBC}
    g_{\bar i} = \frac{ 2 \mathrm{w}_i \gamma k_{\mathrm{r}} C_{\mathrm{eq}}}{\gamma k_{\mathrm{r}} + 2 \mathrm{w}_i \mathbf{e}_{\bar i}\cdot\mathbf{n}} + \frac{-\gamma k_{\mathrm{r}} + 2 \mathrm{w}_i \mathbf{e}_{\bar i}\cdot\mathbf{n} }{\gamma k_{\mathrm{r}} + 2 \mathrm{w}_i \mathbf{e}_{\bar i}\cdot\mathbf{n}} \tilde{g}_i,
\end{equation}
where 
\begin{equation}
    \label{eq:gammaAppendix}
    \gamma := \frac{\tau_{\mathrm{AD}}}{\tau_{\mathrm{AD}} - 0.5} = \frac{1}{1 - \frac{1}{2\tau_{\mathrm{AD}}}}.
\end{equation}

The reformulation of Eqs.~(\ref{eq:PatelBC}) \& (\ref{eq:GuoBC}) are performed with the relation $2 \mathrm{w}_i = c_{\mathrm{s}}^{2}$. It can be applied only for reduced lattices along all directions, except of the rest velocity $i=0$. The latter, however, relates to $g_0$ which is not modified by any of the mentioned boundary schemes.

The reformulation steps required for the scheme of Patel are as follows: Using the second definition of Eq.~(\ref{eq:gammaAppendix}), $\frac{1}{\gamma}$ in Eq.~(\ref{eq:PatelBC}) gives
\begin{equation}
    g_{\bar i} = \frac{ {k_{\mathrm{r}}C_{\mathrm{eq}}-\tilde{g}_{i}\left[{{\frac{k_{\mathrm{r}}}{2\mathrm{w}_{i}}}-\frac{1}{\gamma}\mathbf{e}_{\bar{i}}\cdot\mathbf{n}}\right]}} {{{\frac{k_{\mathrm{r}}}{2\mathrm{w}_{i}}}-\frac{1}{\gamma}\mathbf{e}_{i}\cdot\mathbf{n}}}.
\end{equation}
Multiplying both numerator and denominator with $2\mathrm{w}_i\gamma$ leads to
\begin{equation}
    g_{\bar i} = \frac{ 2\mathrm{w}_i\gamma {k_{\mathrm{r}}C_{\mathrm{eq}}-\tilde{g}_{i}\left[{\gamma k_{\mathrm{r}}-2\mathrm{w}_i \mathbf{e}_{\bar{i}}\cdot\mathbf{n}}\right]}} {{\gamma k_{\mathrm{r}}-2\mathrm{w}_i \mathbf{e}_{i}\cdot\mathbf{n}}}.
\end{equation}
 Using $\mathbf{e}_{\bar{i}} = -\mathbf{e}_{i}$ in the denominator results in
\begin{align}
    g_{\bar i} &= \frac{ 2\mathrm{w}_i {\gamma k_{\mathrm{r}}C_{\mathrm{eq}}-\tilde{g}_{i}\left[{\gamma k_{\mathrm{r}}-2\mathrm{w}_i \mathbf{e}_{\bar{i}}\cdot\mathbf{n}}\right]}} {{\gamma k_{\mathrm{r}}+2\mathrm{w}_i \mathbf{e}_{\bar{i}}\cdot\mathbf{n}}} \\[10pt]
    &= \frac{ 2 \mathrm{w}_i \gamma k_{\mathrm{r}} C_{\mathrm{eq}}}{\gamma k_{\mathrm{r}} + 2 \mathrm{w}_i \mathbf{e}_{\bar i}\cdot\mathbf{n}} + \frac{-\gamma k_{\mathrm{r}} + 2 \mathrm{w}_i \mathbf{e}_{\bar i}\cdot\mathbf{n} }{\gamma k_{\mathrm{r}} + 2 \mathrm{w}_i \mathbf{e}_{\bar i}\cdot\mathbf{n}} \tilde{g}_i,
\end{align}
 which is equal to the scheme of Ju. (cf. Eq.~(\ref{eq:GuoBC})). Thus, the following steps apply to both schemes of Patel and Ju.

The reformulation steps required for the scheme of Ju are as follows: Substitute $2\mathrm{w}_i \mathbf{e}_{\bar i}\cdot\mathbf{n} = c_{\mathrm{s}}^2 \mathbf{e}_{\bar i}\cdot\mathbf{n}$ into Eq.~(\ref{eq:GuoBC}) 
\begin{equation}
\label{eq:reformJu1}
    g_{\bar i} = \frac{ 2 \mathrm{w}_i \gamma k_{\mathrm{r}} C_{\mathrm{eq}}}{ \gamma k_{\mathrm{r}} + c_{\mathrm{s}}^2 \mathbf{e}_{\bar{i}}\cdot\mathbf{n}} + \frac{-\gamma k_{\mathrm{r}} + c_{\mathrm{s}}^2 \mathbf{e}_{\bar{i}}\cdot\mathbf{n} }{\gamma k_{\mathrm{r}} + c_{\mathrm{s}}^2 \mathbf{e}_{\bar{i}}\cdot\mathbf{n}} \tilde{g}_i.
\end{equation}
Then divide both numerator and denominator by $c_{\mathrm{s}}^2 \mathbf{e}_{\bar{i}}\cdot\mathbf{n}$.
\begin{align}
\label{eq:reformJu2}
    g_{\bar i} &= \frac{ 2 \mathrm{w}_i\ \frac{\gamma k_{\mathrm{r}}}{c_{\mathrm{s}}^2 \mathbf{e}_{\bar{i}}\cdot\mathbf{n}}\ C_{\mathrm{eq}}}{ \frac{\gamma k_{\mathrm{r}}}{c_{\mathrm{s}}^2 \mathbf{e}_{\bar{i}}\cdot\mathbf{n}} + 1} + \frac{-\frac{\gamma k_{\mathrm{r}}}{c_{\mathrm{s}}^2 \mathbf{e}_{\bar{i}}\cdot\mathbf{n}} + 1 }{\frac{\gamma k_{\mathrm{r}}}{c_{\mathrm{s}}^2 \mathbf{e}_{\bar{i}}\cdot\mathbf{n}} + 1} \tilde{g}_i \\[10pt]
\label{eq:reformJu3}
    &= \frac{ 2 \mathrm{w}_i\ k_{i}^{(\mathrm{P,J})}\ C_{\mathrm{eq}}}{ k_{i}^{(\mathrm{P,J})} + 1} + \frac{-k_{i}^{(\mathrm{P,J})} + 1 }{k_{i}^{(\mathrm{P,J})} + 1} \tilde{g}_i 
\end{align}
The final equation is the same as of Verhaeghe. (cf. Eq.~(\ref{eq:fluxBCSolve})) however with a different definition of $ k_{i}^{(\mathrm{S.P.J.})} = c_{\mathrm{s}}^{-2} \gamma k_{\mathrm{r}} /\ (\mathbf{e}_{\bar i}\cdot\mathbf{n})$. 

To summarize, after reformulating the schemes the following general scheme is the result
\begin{equation}
    g_{\bar i} = \frac{k_i}{1+k_i} 2 \mathrm{w}_i C_{\mathrm{eq}} + \frac{1-k_i}{1+k_i} \tilde{g}_i.
\end{equation}
The difference of the schemes only enters by the definition of the term $k_i$
\begin{align}
     k_{i}^{(\mathrm{S.V.})} &= c_{\mathrm{s}}^{-2} \ \  k_{\mathrm{r}} \cdot (\mathbf{e}_{\bar i}\cdot\mathbf{n}), \\
     k_{i}^{(\mathrm{S.P.J.})} &= c_{\mathrm{s}}^{-2} \gamma k_{\mathrm{r}} /\ (\mathbf{e}_{\bar i}\cdot\mathbf{n}), \\
     k_{i}^{(\mathrm{S.New})} &= c_{\mathrm{s}}^{-2} \gamma k_{\mathrm{r}} \cdot (\mathbf{e}_{\bar i}\cdot\mathbf{n}).
\end{align}
%}

% \vspace{2em}\noindent\textbf{Text S1.3. Behavior at PeDa$\rightarrow\infty$ limit } 
\subsection{Behavior at PeDa$\rightarrow\infty$ limit}
\label{sec:Appendix_ResultsSchemePeDaLimit}
%{
At $k_{\mathrm{r}}\rightarrow\infty$ limit, i.e.\, Da\;$\rightarrow\infty$, the RBC schemes approach the anti bounce-back (anti-BB) scheme \citeA{Verhaeghe2006, Ju2020, Patel2016}. This holds also for PeDa\;$\rightarrow\infty$ (cf. Section~3.1.3 in the main paper) as is shown in the following where only the Pe number is varied. % \ref{sec:ResultsSchemePeDaLimit}

Regarding the outgoing population, the difference between the anti-BB (ABB) and the RBC scheme is
\begin{equation}
    \label{eq:ABBvsGBC}
    \Delta g_{\bar{i}}^{\ (\mathrm{ABB\, vs\, RBC})} \approx \frac{2}{1+k_{\bar{i}}}\left( \mathrm{w}_i C_{\mathrm{w}} - \tilde{g}_i \right).
\end{equation}
This is derived as follows; starting with
\begin{align}
    g_{\bar{i}}^{\ (\mathrm{ABB})} &= 2\mathrm{w}_i C_{\mathrm{w}} - g_{i}^{\ (\mathrm{ABB})} \\
    g_{\bar{i}}^{\ (\mathrm{RBC})} &= \frac{k_{\bar{i}}}{1 + k_{\bar{i}}} 2 \mathrm{w}_i C_{\mathrm{eq}} +  \frac{ 1 - k_{\bar{i}}}{1 + k_{\bar{i}}} g_{i}^{\ (\mathrm{RBC})} ,
\end{align}
the difference is
\begin{align}
    \Delta g_{\bar{i}} &= \left| g_{\bar{i}}^{\ (\mathrm{ABB})} - g_{\bar{i}}^{\ (\mathrm{RBC})} \right| \\[10pt]
    &= 2\mathrm{w}_i C_{\mathrm{w}} - g_{i}^{\ (\mathrm{ABB})} - \left( \frac{k_{\bar{i}}}{1 + k_{\bar{i}}} 2 \mathrm{w}_i C_{\mathrm{eq}} +  \frac{ 1 - k_{\bar{i}}}{1 + k_{\bar{i}}} g_{i}^{\ (\mathrm{RBC})} \right) \\[10pt]
    &= \frac{2\mathrm{w}_i}{1+k_{\bar{i}}} ( C_{\mathrm{w}} + k_i (C_{\mathrm{w}} - C_{\mathrm{eq}}) ) - g_{i}^{\ (\mathrm{ABB})} - \frac{ 1 - k_{\bar{i}}}{1 + k_{\bar{i}}} g_{i}^{\ (\mathrm{RBC})}
\end{align}
Here, it is assumed that cells in direct vicinity to a boundary cell have approximately the same concentration for large PeDa values, i.e.\ $\tilde{g}_{i}^{\ (\mathrm{ABB})} \approx \tilde{g}_{i}^{\ (\mathrm{RBC})}$ and $C_{\mathrm{w}} \approx C_{\mathrm{eq}}$. 
\begin{align}
    \Delta g_{\bar{i}} &= \frac{2\mathrm{w}_i C_{\mathrm{w}}}{1+k_{\bar{i}}} - \frac{2}{1 + k_{\bar{i}}} g_{i} \\[10pt]
    &= \frac{2}{1+k_{\bar{i}}}\left( \mathrm{w}_i C_{\mathrm{w}} - g_i \right)
\end{align}
Here, the $\mathrm{w}_i C_{\mathrm{w}} - g_i$ term is the discrepancy between the incoming and outgoing population, which is indicative of the gradient at the boundary itself.

The RBC schemes to approach the anti-BB scheme, i.e.\, $\Delta g_{\bar{i}}^{\ (\mathrm{ABB\, vs\, RBC})}  \rightarrow 0 $, under the following conditions. Either for $2\ /\ (1+k_{\bar{i}})\rightarrow0$ which holds for increasing values of $k_{\bar{i}}$ and Da, or for $\left( \mathrm{w}_i C_{\mathrm{w}} - \tilde{g}_i \right)\rightarrow0$ which holds for a decreasing gradient of the wall concentration and increasing value of Pe. Finally, both conditions together show that the RBC schemes approach the anti-BB scheme as PeDa\;$\rightarrow\infty$.
%}

\section{Detailed error analysis of the rest fraction method} \label{sec:Appendix_restFraction}
%{
The additional error introduced by applying the rest fraction method is quantified using the MAE. Simulations are conducted using the new scheme for the simulation setup of configuration~A of the 2D reaction-diffusion problem described in Section~3.1. % \ref{sec:resultsRBCComparison}
It is tested against the simulation without rest fraction as baseline. Both BGK and TRT collision operators are used with the following parameter variations: $J_0= \{0.1, 0.\bar{3}, 0.5, 0.9\}$ at PeDa=1, where $k_{\mathrm{r}}$ is kept constant resulting in $\tau_{\mathrm{AD}}=\{0.6\bar7, 0.74, 0.82, 2.1\}$. The results are shown in Table \ref{tab:RestFractionResults}.  

\begin{table}[ht!]
\centering
\caption{\textit{Impact of the Rest Fraction Method on Configuration~A Simulations}}
\label{tab:RestFractionResults}
\begin{tabular}{ r | c c c c | l}
            & \multicolumn{4}{c|}{$J_0$} & Baseline \\
            & 0.1 & 0.$\bar{3}$ & 0.5 & 0.9 & \\ \hline 
 $c_{\mathrm{s}}^2$ & 0.45 & $0.3\bar{3}$ & 0.25 & 0.05 & $0.3\bar{3}$ \\
% $\tau_{\mathrm{AD}}$         & 0.6$\bar7$  & 0.74   & 0.82   & 2.1 & 0.74   \\
 MAE BGK $\cdot 10^{-3}$ & 1.212 & 1.213 & 1.214 & --- & 1.213 \\
 MAE TRT $\cdot 10^{-3}$ & 1.219 & 1.218 & 1.217 & 1.200 & 1.218
\end{tabular}
\end{table}

As expected results for the $J_0 = 0.\bar{3}$ and the baseline are identical. For the BGK simulations, the $J_0=0.9$ simulation was unstable due to a too high $\tau_{\mathrm{AD}}$, which can be mitigated with the TRT collision operator. For both BGK and TRT collision operators, the difference in the MAE between the baseline and the rest fraction simulations are negligibly higher. Overall, the rest fraction method and TRT collision operator can be used to simulate large Pe to Re ratios.

\newpage
% \vspace{2em}\noindent\textbf{Text S2. Reactive Object in Channel Flow} 
\section{Reactive Object in Channel Flow}
\label{sec:Appendix_reactiveObject}
%{
Molins~{et al.}'s paper uses five different simulation methods (Chombo-Crunch, dissolFoam, LBM, OpenFoam and Vortex) to simulate the average reaction rate $R_{\mathrm{avg}}$ of a reactive object in channel flow. The exact method of how these results were aggregated is shown here. Only the minimum and maximum values were used in Table~2 in the main paper. %\ref{tab:MolinsVerification}

The results of Molins~{et al.} are summarized in Table~\ref{tab:MolinsAggregation}.
For the dissolution cases 2), 3) and 4) the quasi-steady reaction rates are given. They are extracted from their supplementary information (SI), considering the up to five timesteps before dissolution, i.e.\ while the reported object's surface area is unchanged.
The number of timesteps used is reported in brackets, e.g.\ dissolFoam (2) has two time steps. This should give an indication on the object's $R_{\mathrm{avg}}$ behavior and order of magnitude.
\begin{table}[!htbp]
\centering
\caption{\textit{Result of $R_{\mathrm{avg}}$ of Different Simulation Methods from Molins~{et al.}}}
\label{tab:MolinsAggregation}
\begin{tabular}{c r l | l l l l l }
\multicolumn{3}{c|}{Input} &  \multicolumn{5}{c}{$R_{\mathrm{avg}}$ $[10^{-8}\; \mathrm{mol/(cm^2\cdot s)}]$} \\
Case & Pe & Da & Chombo-  & dissolFoam & LBM & OpenFoam & Vortex \\[-5pt]
\multicolumn{3}{r|}{} & Crunch (5) & (2) & (3) & (2) & (5)   \\ \hline
1) & 600 & 178   & $4.32$ & $4.33$ & $4.57$ & $4.18$ & $4.27$ \\
2) & 600 & 17800 & $4.58$ & $4.54$ & $4.79$ & $3.88$ & $4.71$ \\
3) & 6   & 178   & $58.6$ & $75.8$ & $75.3$ & $74.2$ & $90.5$ \\
4) & 6   & 0.178 & $7.51$ & $8.06$ & $8.04$ & $ - $ & $ - $ \\
\end{tabular}
\end{table}

Noteworthy is that in their SI, the Vortex results states it simulated at Re\;$=0.24$ and Chombo-Crunch uses Re\;$=1.0$ as input. This may explain the discrepancies in case 3). For case 4) only data of three of the five are available, of which one is Chombo-Crunch with Re\;$=1.0$. This leaves two simulation results: dissolFoam and LBM. The LBM simulation may not have the RBC with correct wall diffusion, thus it makes sense that the scheme of Verhaeghe agrees with it. Remaining is one simulation method (dissolFoam) consisting of two data points, making it difficult to draw accurate conclusions about the accuracy of the new scheme. 

\newpage

\begin{figure}[ht!]
    \centering
    \includegraphics[width=\linewidth]{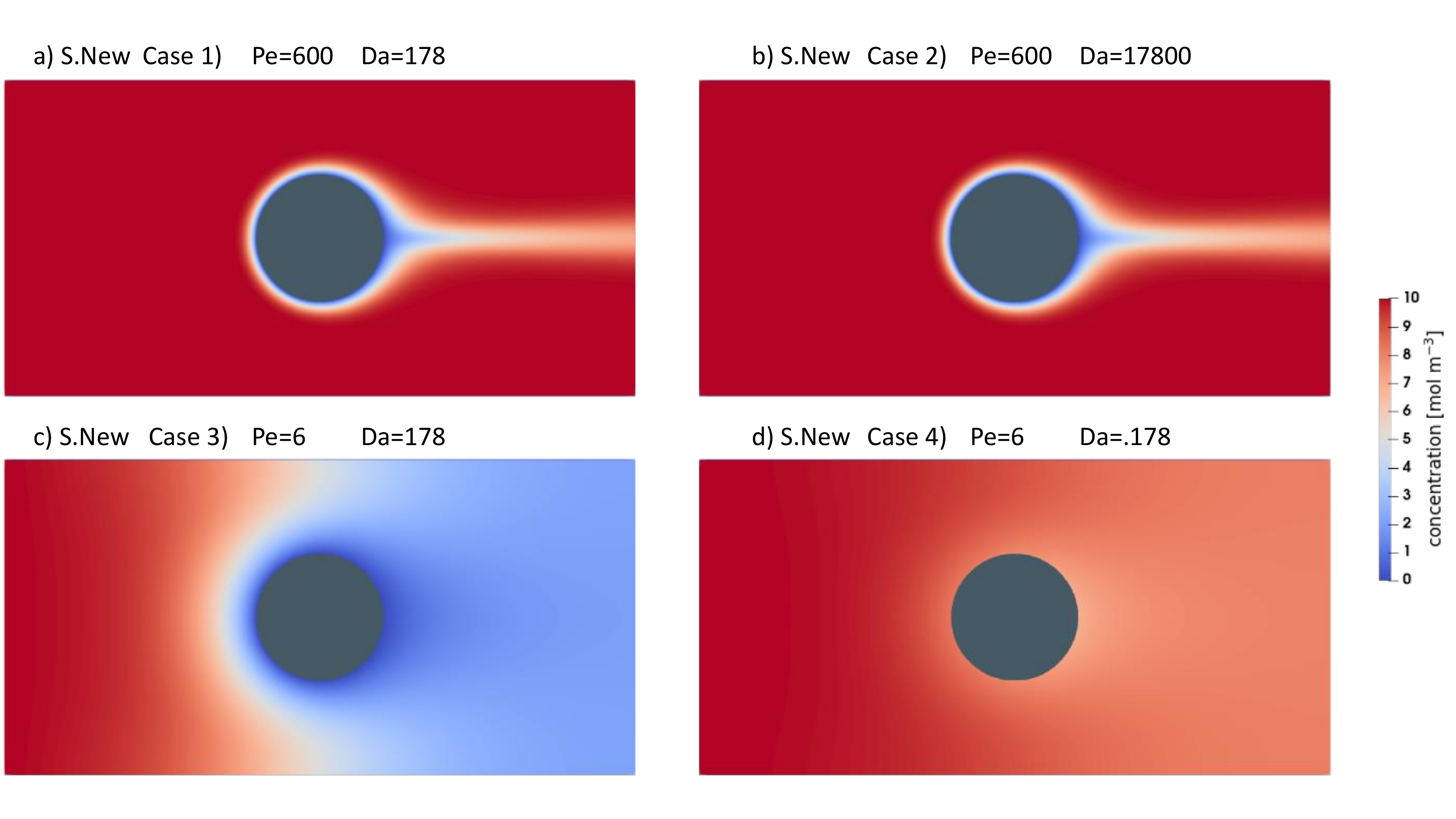}
    \caption{Concentration field in the vicinity of a reactive object (grey circle) in a channel flow simulated using the new scheme (cf. 3.3 \textit{Reaction in a Channel Flow} in the main paper). a) \& b) Cases 1) \& 2) strongly diffusion limited and advection-reaction driven, causing the tail with low concentration behind the object. c) Case 3) advection-diffusion limited and reaction driven, indicated by a broader area of low concentration. d) Case 4) advection-reaction-diffusion limited, with high concentration at the object's surface. The concentration is shown by the color code given in the legend. }
    \label{fig:molinsResultsS3}
\end{figure}

\bibliography{ms}